\documentclass[10pt]{article}
\usepackage{graphicx}
\newdimen\mypicsize
\mypicsize=\hsize
\title{The Static Universe Hypothesis:\\
Theoretical Basis\\ and \\
Observational Tests of the Hypothesis}

\author{Thomas B. ANDREWS\\(tom-andrews@msn.com)}

\date{September 6, 2001}

\newcommand{\pder}[2]{\frac{\partial #1}{\partial #2}}
\newcommand{\pdersq}[2]{\frac{\partial^{2}#1}{\partial #2^{2}}}

\begin{document}

\maketitle

\begin{abstract}

From the axiom of the unrestricted repeatability of all experiments,
Bondi and Gold (also Hoyle) argued that the universe is in a stable,
self-perpetuating equilibrium state. Their reasoning extends the usual
cosmological principle to the perfect cosmological principle in which
the universe looks the same from any location at any time. By itself,
the perfect cosmological principle predicts the universe is static and
in an equilibrium state.

However, Bondi and Gold rejected the static universe prediction for two reasons: First,
they believed the universe was expanding because of the Hubble redshift. Second, the universe
appeared to be far from thermodynamic equilibrium. Therefore, they hypothesized that the
universe was expanding and in a steady-state. The steady-state universe is an expanding
universe model in which matter is created. As the galaxies recede, new galaxies form from the
created matter and maintain the universe in a stationary state.

Instead of the steady-state model or the current Friedmann-Walker expanding universe model,
I hypothesize that the universe is static and in an equilibrium state (non-evolving)
as predicted by the perfect cosmological principle.

New physics is proposed based on the concept that the universe is a pure wave system.
Based on the new physics and assuming a static universe, new processes are derived for the
Hubble redshift and the cosmic background radiation field.

A new time-dilation process is proposed as the cause of the anomalous dimming of Type
Ia supernovae at high $z$. This process is based on the Hubble redshift
increasing the period of the supernovae luminosity curve in the
observer's rest frame. In turn, the increase in the period reduces the
luminosity of supernovae by $1/(1 + z)$. Furthermore, since this
process is independent of the cause of the Hubble redshift, the new
time-dilation process must also apply to supernovae in the expanding
universe models. But, the expanding universe model already incorporates
a time-dilation effect which applies to any object. Thus, two
time-dilation effects should be observed for supernovae. Since only one
time-dilation effect is observed for supernovae, the expanding universe
model is logically falsified.

Following the scientific method, I test deductions developed from the static
universe hypothesis using observational data primarily from the Hubble Space Telescope.
Applying four different global tests of the space-time metric, I find that the
observational data consistently fits the static universe model and, therefore,
confirms the static universe hypothesis. The observational data also show that the
average absolute magnitudes and physical radii of first-rank elliptical galaxies
have not changed over the last $5$ to $10$ billion years, thereby confirming the
perfect cosmological principle.

In the expanding universe models, the observed baryonic mass density is
a factor of $25$ to $50$ lower than the predicted mass density in a
flat universe. This discrepancy between theory and observation is a
major problem and has resulted in many hypotheses concerning the nature
of the ``missing mass.'' In the static universe model, the predicted
baryonic mass density is lower by a factor of about $20$. Consequently,
the discrepancy in the baryonic mass density between theory and
observation is removed.

Because the static universe hypothesis is a logical deduction from the perfect
cosmological principle and the hypothesis is confirmed by the observational data,
I conclude that the universe is static and in an equilibrium state.

\end{abstract}

\pagebreak

\tableofcontents

\pagebreak

\section{Introduction}
\label{sec:INTR}

The current standard model of the universe is the Friedmann-Walker
expanding universe. The major  factors which led to the adoption of the
expanding universe model in the late 1920's and early 1930's were:

\begin{enumerate}
  \item The distribution of the galaxies is homogeneous and isotropic.
  \item The galaxies are receding from us with velocities proportional to their distances.
\end{enumerate}
The first factor is undoubtedly correct. Observations show that the galaxies on a large
enough scale are distributed homogeneously and isotropically. However, the recession of
the galaxies was based on the Hubble redshift which remains an assumption. Initially, the
redshift was assumed due to the Doppler shift process and, later, to the expansion of space.
Thus, the Hubble redshift could still be due to a different process.

If another process is responsible for the Hubble redshift, the theoretical picture becomes
quite different. The universe would be static rather than expanding. This, of course, is
only a conjecture at this point but it is an important first step.

A few cosmologists, notably Jaakkola~\cite{GR:JK} and LaViolette~\cite{GR:LV}, have proposed
that the universe is static. Jaakkola has argued that the universe is static and in an equilibrium state. LaViolette has shown that the observational data of the mid-1980's is more consistent with the static universe model than the expanding universe models and proposed a new process for the Hubble redshift. I also proposed~\cite{A:g} that the universe was static in 1994. However, at that time, the observational data was not good enough to convincingly prove that the universe was static. On the other hand, Sandage and Lubin~\cite{ST:LS} have recently concluded from surface brightness observations of elliptical galaxies that the universe is expanding.

This paper is equally divided between theory and observation. The theoretical part is required
to derive the Hubble redshift and predict the mass density of the universe. The observational
section is also large in order to adequately describe and analyze the observational data sets.
Consequently, the paper is organized as follows:

\begin{enumerate}
 \item The static universe hypothesis is developed. This is an important first step
  in the scientific method~\cite{ST:ROM}.
 \item New physics is introduced in order to derive the Hubble redshift. New physics
  in this instance appears required since all previous attempts to derive the Hubble
  redshift using current physics appear to have failed.
 \item The source of the mass-energy of elementary particles is determined. In particular, an
  average mass density of the universe consistent with observation is predicted.
 \item The size and physical nature of the universe are discussed. This section contains
  another argument for a static and equilibrium universe.
 \item A new Hubble redshift process is derived for a static universe. This process applies to
  both photons and mass particles.
 \item A theoretical solution to the anomalous dimming of supernovae is proposed based on a
  new process involving time-dilation of the supernovae light curve.
 \item An equilibrium energy process is proposed for the Cosmic Microwave Background (CMB)
  radiation field in a static universe.
 \item The deductions of the static universe hypothesis are compared to the
  observations, using four different global tests of the space-time metric.
\end{enumerate}

\section{The Static Universe Hypothesis}
\label{sec:SUH}

Rather than argue for a static universe hypothesis as in the
introduction, it is better to develop the hypothesis based on physical
logic. Thus, the hypothesis of a static universe in an equilibrium
state follows directly from the perfect cosmological principle (PCP)
which was proposed in 1948 by Bondi and Gold~\cite{GR:BG} (also
Hoyle~\cite{ST:FH}). The PCP says the universe looks the same from any
location at any time.

Bondi and Gold based the reality of the PCP on the following fundamental
assumption:
\begin{quote}
  Given that the unrestricted repeatability of all experiments is a
  fundamental axiom of physical science, this implies that the outcome of an
  experiment is not affected by the location and time at which it is
  carried out.
\end{quote}
They believed that cosmology must be concerned with this fundamental assumption and, in turn,
an adopted cosmology must incorporate this assumption.

Based upon this fundamental assumption, Bondi and Gold then developed the following paradigm:
\begin{quote}
  As the physical laws cannot be assumed independent of the structure of
  the universe and, conversely, the structure of the universe depends
  upon the physical laws, it follows that the universe is in a
  stable self-perpetuating state, without making any assumptions regarding
  the particular features which lead to this stability.
\end{quote}
They emphasized that only in such an equilibrium universe can the constants
and laws of physics be invariant to both changes in location and time.

Of course, Bondi and Gold did not absolutely claim that the PCP must be
true. However, they believed that if it does not hold,
the variability of the physical laws becomes so wide that one can no
longer use local physics in the distant universe without relying on
arbitrary principles for the extrapolation of local physics.

Conversely, if the PCP holds in the universe, we can confidently base
our results on the permanent validity of all our experiments and
observations. Therefore, they concluded that we should proceed
theoretically assuming the PCP is true since this is the only basis
permitting progress without further arbitrary assumptions.

At the present time, the PCP must be considered an even stronger
theoretical principle since the invariance of the
physical constants~\cite{GR:VV} and the laws of physics are confirmed both by local
experiments and by distant observations of the universe.

From the PCP, Bondi and Gold at first expected the universe to be static.
However, because of the observed redshifts of the galaxies and also the active state of
universe, they believed the universe must be expanding and thermodynamically
in a non-equilibrium state.  Then, in order to maintain a stationary state of the
universe even though the universe was expanding, they were forced to
assume that matter is created in the voids left by the expansion of the
space between the galaxies. They referred to this new model of the universe as
the steady state universe.

From today's perspective, I believe their initial expectation that the
universe was static was correct but that their further assumptions
leading to an expanding universe with matter creation were both
unnecessary and incorrect. However, in 1948 when they developed the PCP, the
steady state universe was possibly the only logical way to proceed.

Currently, based on the derivation of a new process for the Hubble
redshift in a static universe, many of the elements of a static universe
are understood. In any case, the static universe hypothesis can now be tested
since precision observational data is now available from the Hubble Space
Telescope and the new, large ground telescopes.

\section{New Physics}
\label{sec:NPH}

The basic problem is to develop a new process for the Hubble redshift in
a static universe. Since the Hubble redshift has been known for over 70 years, a large number
of attempts have been made to derive a physical process for the Hubble redshift in a
static universe. The requirements for the Hubble redshift process are:

\begin{enumerate}
  \item The shift in frequency is strictly proportional to the frequency.
  \item All electromagnetic frequencies are equally affected.
  \item The redshift is proportional to the distance.
\end{enumerate}
The first two requirements are satisfied by the Doppler process but, of
course, it is not the Doppler process which is now held responsible for
the redshift but the expansion of the universe. The expansion fits all
of the above requirements.

Gravitation is the only other known process with the same properties. So, without
introducing a new force, gravitation is an obvious candidate for the
Hubble redshift. However, in current physics, gravitation is too weak a force to cause
the Hubble redshift. This is the current dilemma if new physics is not introduced.

\subsection{The Wave System Theory}
\label{sec:WST}

As a more fundamental and comprehensive viewpoint in physics, I propose {\em the Universe is a Pure Wave System consisting of a large number of wave modes.} This is a new paradigm that was inferred from the universal occurrence of wave phenomena in physics and the consistency of the paradigm with the basic laws of physics. From this concept of a wave system, the existence of particles, fields and quantum effects~\cite{A:quant} may be derived.

Because the wave modes have very small amplitudes, I
assume the wave modes can be modeled using the classical linear wave equation
  \begin{equation}
    \pder{}{x}\left(T(x)\pder{\phi}{x}\right)
    = \sigma(x)\pdersq{\phi}{t}
    \label{eq:1}
  \end{equation}
where the parameters, $T(x)$ and $\sigma(x)$, are the tension and (linear) mass density
respectively. The parameters are not fixed but can vary subject to local constraints. And,
because the parameters can vary, the wave modes can exchange energy through parametric
interactions.

The wave modes can be shown to constructively interfere and produce localized peaks surrounded
by large regions of destructive interference. It is then hypothesized that these
localized peaks are the elementary particles. As derived later, the wave system becomes a
deeply bound system when the localized peaks are formed. The system then appears to consist
only of stable, interacting constructive interference peaks, indistinguishable in
properties from the observed elementary particles.

A symmetry argument of H. Giorgi~\cite{GR:HG}~proves that the wave system exists. Given a
system which is infinite and linear and where the laws of physics are space and time translation invariant, Giorgi argues that the modes of oscillation of a system are determined by the representations of the space and time translation groups. Since a solution of each representation is a complex exponential in space and time respectively, a standing wave system is formed, given by $\exp{i(\omega t \pm kx)}$.

The symmetry proof must apply to the universe since the properties of the universe match very
closely the requirements for the symmetry. First, the universe is a very large if not infinite
system with linear responses at the level of basic laws. And second, the exact time and
translational invariance of the laws of nature has been experimentally confirmed.

C. Vassallo~\cite{GR:CV} reaches similar conclusions. He also finds that fields which are
invariant to space and time translation must vary as $\exp{i(\omega t \pm kx)}$. In addition,
he finds that any bounded wave field which has sources can be represented by normal modes
which do not have sources. This is a significant theorem. It suggests that a system of
bounded normal modes can bootstrap initial sources to produce the normal modes.

The stability of the wave system is based on the following two
principles:

\begin{enumerate}
  \item The frequency is reduced if the mass density and
  tension are larger at the wave mode peaks.
  \item As the number of elementary particles mutually interacting
  increases, the frequency increases.
\end{enumerate}
To apply the first principle, assume the mass density and tension are
proportional to the local energy density of the wave system. Then, the
frequency of a wave mode will decrease when the constructive
interference of the wave modes produces high energy concentrations at
the peaks of the wave mode (See equation~\ref{eq:6L} and accompanying
discussion). As an example of this principle, consider a string
vibrating in it's lowest frequency mode. When lead weights are placed
at the peaks of the vibration, the frequency is reduced. Similarly,
when the tension is increased at the peaks, the frequency is reduced.
Since the energy density at a peak is proportional to the number of
wave modes $(N \approx 10^{38})$ constructively interfering, the
decrease in frequency of the wave modes is very large. If the natural
frequency of the wave system without constructive interference is
$f_o$, complete constructive interference reduces the frequency to
\begin{equation}
  f_m = \frac{f_o}{\sqrt{N}}.
  \label{eq:N}
\end{equation}

Since the number of wave modes increases as the average distance
between particles increases, the frequency could go to zero as $N$
increases without limit. However, by the second principle, there is a
lower bound on the frequency determined by the number of interacting
particles in the universe. This frequency is given by the simple
eigenvalue equation proposed by Chen~\cite{GR:FC}
  \begin{equation}
     f_p = \frac{1}{2\pi}\sqrt{n} \sqrt{k/m}
     \label{eq:chen}
  \end{equation}
where $n$ is the effective number of particles interacting and the interaction
constant, $k/m$,  between any two particles is constant.

This eigenvalue system is remarkable since it has only two discrete frequencies, one
degenerate with $n-1$ modes given by $f_p$ and the other equal to $f_o = f_p/\sqrt{n}$,
the frequency of a single element and also the minimum frequency. Equation~\ref{eq:chen}
is the approximate eigenvalue solution to the interactions between all the
particles $(\approx 10^{80})$ in the universe. This assumes that in the universe the
interactions between particles have the same average strength.

A stable frequency of the wave system is attained when the wave mode frequency, $f_m$, and
the particle frequency, $f_p$, are the same. The particle frequency may be estimated from
equation~\ref{eq:chen}. Let $f_o = 1/(2\pi)\sqrt{k/m} = c/(2R)$. $R$ is the mean absorption
distance and is equal to $c/H$ (see section~\ref{sec:HRP}). $c/(2R)$ estimates the minimum
frequency, $f_o$. Then, assuming $R = 1.85 \times 10^{28}$ cm (H = 50 km/sec/Mpc)
and $n \approx 10^{80}$ particles, the stable frequency is $8.1 \times 10^{21}$ Hertz.

As a first approximation, the wave system vibrates at this single frequency. In the second
approximation, the single frequency is split into two frequencies by the interactions
between the particles. The higher frequency is hypothesized to correspond to the proton
and the lower frequency to the electron. This eigenvalue theory is consistent with the fact
that these are the only stable mass particles.

The number of normal modes is an important parameter in the wave system theory because the
properties of the wave system depend on constructive interference. In particular, the ratio
between the electrostatic force and the gravitational force is proportional to the number of
modes interfering. This can be shown as follows: For constructive interference, the
intensity, $I$, at a particle is proportional to
 \begin{equation}
   I \propto \sum{\left(A_1 + A_2~...~A_N \right)^2} = NA^2 \pm 2 N^2A^2
   \label{N:1}
 \end{equation}
where $A$ is the amplitude of each mode of vibration. It is assumed that the force of
gravitation is due to the $NA^2$ terms since gravitation is a small effect which is always
attractive. On the other hand, the electrostatic force is assumed proportional to the $N^2A^2$
terms. For $N$ on the order of $10^{38}$, the electrostatic force is $10^{38}$ times larger
than the gravitational force and can be either plus or minus.

It would be natural to assume that the intensity of the wave modes at a particle is primarily
due to the interference terms. However, the intensity at a particle is given by
 \begin{equation}
   I \propto NA^2
 \end{equation}
since the $2N^2A^2$ terms in equation~\ref{N:1} cancel on the average at an elementary
particle, This is understandable because the electrostatic energy is equally plus and
minus at a particle. Since only the squared terms remain, the mass energy of elementary
particles is then purely gravitational in origin rather than overwhelmingly electrostatic.

\subsection{General Force Equation}

In order to determine the forces which occur in the wave system, the wave system
equation must be solved for small changes in the variable parameters. This assumes
that all forces are due to changes in the parameters.

Since the equilibrium solution of the wave system equation is needed, the
space dependent eigenvalue equation will be used. This is derived from equation~\ref{eq:1}
by separation of variables and is
  \begin{equation}
    \frac{d}{dx}\left(T\left(x\right) \frac{dY}{dx}\right)
    + \sigma\left(x\right) (2 \pi f)^{2}Y = 0
    \label{eq:3}
  \end{equation}
where $(2 \pi f)^2$ is the separation constant.

For small spatial variations in the parameters, the perturbative solution of
equation~\ref{eq:3} is given by
\begin{eqnarray}
   f^{2} & = & f_o^{2}\left(1 - \frac{2}{L\sigma_{o}}
    \int_{0}^{L}\left(\sigma(x) - \sigma_{o}\right)\sin^{2}\left(kx\right)dx \right. \\
         & - & \left.\frac{2}{kLT_{o}} \int_{o}^{L} {\pder{T(x)}{x}
               \cos\left(kx\right) \sin\left(kx\right) dx} \right) \nonumber
  \label{eq:57}
\end{eqnarray}
where \mbox{$f_{o} = k^{2}_{o}/(2 \pi)^2 \left(T_{o}/
\sigma_{o}\right)$} is the unperturbed frequency,
$\sigma_{o}$ and $T_{o}$ the average mass density and tension and
$L$ is the size of the system. This equation shows that $f$
decreases when $\sigma(x)$ and $\partial{T(x)}/\partial{x}$ are larger
at the constructive interference peaks.

Equation~\ref{eq:57} may be simplified as follows: First, by noting
that the last two terms are equal. Second, by setting $\sin^2{(kx)}$ and
$\sin{(kx)} \cos{(kx)} = 1/2$ and taking the square root. Then, the perturbed frequency is
approximated by
\begin{equation}
    f = f_{o} \left(1-\frac{1}{L\sigma_{o}}\int_{0}^{L}
    {\left(\sigma(x) - \sigma_{o}\right)dx} \right).
    \label{eq:5L}
  \end{equation}

Now assume there is only a single particle in the system and the particle or constructive
interference peak has a linear width equal to the wavelength, $\lambda$.
To simplify the nomenclature, set $\sigma = \sigma(x)$. Then, $\sigma = m/\lambda$
and $\sigma_{o} = m/L$ where $m$ is the mass of the particle. Integrating
equation~\ref{eq:5L},
  \begin{equation}
     f = f_o \left(2 - \frac{\lambda \sigma}{L \sigma_{o}}\right).
     \label{eq:6L}
  \end{equation}
Note: $\sigma dx$ integrates to $\lambda \sigma$ since $\sigma$ is a constructive interference
peak which only exists over one wavelength within the much larger integration distance, $L$.
Without the constructive interference peak, $f = 2f_o$. With the constructive interference
peak, $f = f_o$, the stable frequency of the wave system.

Setting $E= hf$, $mc^2= hf_{o}$ and $m = L \sigma_o$ in equation~\ref{eq:6L}, we have
  \begin{eqnarray}
     E & = & mc^2 \left(2 - \frac{\lambda \sigma}{L \sigma_o} \right) = 2mc^2 - c^2
     \lambda \sigma \\
       & = & 2mc^2 -  \lambda \sigma_e  \nonumber
         \label{eq:4P}
  \end{eqnarray}
where $\sigma_e$ is the energy density (still a function of x).
Then the force required to move a mass particle is given by
  \begin{equation}
      F = \frac{dE}{dx} = - \lambda \frac{d \sigma_e}{dx}.
         \label{eq:5P}
  \end{equation}
Equation~\ref{eq:5P} is the ``general force equation.'' The type of force depends upon the
nature of the physical process which changes $\sigma_e$.

\subsection{Derivation of Newton's Law of Gravitation}\label{sec:4.0}

In a previous paper~\cite{GR:TA}, I derived Newton's law of gravitation
from the wave system theory. Because understanding gravitation on a
deeper level than Newtonian gravitation or, for that matter, general
relativity is essential to the cosmological theories proposed in this
paper, I am repeating the derivation in this paper.

To derive Newton's law, assume two protons, labeled $m_1$ and $m_2$, are $r$ cm apart.
The model for gravitation is as follows: A wave mode originating at $m_1$ interacts with
$m_2$. Quantitatively, the gravitational force can be calculated from equation~\ref{eq:5P}.
The first step is to derive the energy density of a single wave mode at $m_2$ as a function
of $r$. It is
assumed that the initial energy density, $\sigma_e$, of a single normal mode is $1/N$ times
the energy density at a proton where $N$ is the number of normal modes. Furthermore, the
natural assumption in three-dimensional space is that $\sigma_e$ decreases as $1/r^2$.
But differentiating $\sigma_e$ with respect to $r$, the force varies as $1/r^3$. This is
obviously an incorrect result since the gravitational force experimentally varies as $1/r^2$!
But why is this incorrect result obtained?

Assuming the derivation of the general force law is valid, I was forced
to reject the isotropic propagation of the normal modes. Instead, I
assumed that each wave mode propagates circularly in a plane. The
energy density then varies as $1/r$. Although the circular propagation
of the normal modes in planes appears physically improbable, I found
that the assumption worked perfectly.

Applying this assumption, the energy density of a single normal mode is
\begin{equation}
    \sigma_e = \frac{m_1 c^2}{2\pi N r}~\hbox{erg/cm}.
    \label{eq:26M}
  \end{equation}
Then, from the General Force Law equation~\ref{eq:5P}, the gravitational force is
  \begin{equation}
    F = - \lambda \frac{d\sigma_e }{dr} = 
	\lambda \frac{m_1c^2}{2 \pi N r^2}~\hbox{dynes}
    \label{eq:27M}
  \end{equation}
where $\lambda$ is the linear size of an elementary particle, i.e., for a proton $\lambda
= 1.3 \times 10^{-13}$ cm.

The dependence of the gravitational force law on the size of the particle is eliminated by
the following principle: {\em The size of a particle is proportional to it's mass.} This
follows since the mass density must be the same for all particles; otherwise, energy would be
transferred between particles to reach a lower system eigenvalue. We can, therefore, introduce
an equilibrium mass density constant for all particles,
  \begin{equation}
    \sigma_m  =  \frac{m_p}{\lambda} = 1.28 \times 10^{-11}~\hbox{g/cm}
    \label{eq:57M}
  \end{equation}
where the mass of a proton is $1.67 \times 10^{-24}$ g. The constant, $\sigma_m$, explains
why the electron is so much smaller than the proton.

To obtain the exact form of Newton's law, multiply equation~\ref{eq:27M} by $m_2/m_2$ and set
 \begin{equation}
  G = \frac{c^2}{2 \pi N \sigma_m}.
  \label{eq:28M}
 \end{equation}
Then, the correct form of Newton's law of gravitation is obtained,
  \begin{equation}
    F = G \frac{m_1 m_2}{r^2}.
  \end{equation}
This equation has been explicitly derived only for the force between two protons. However,
since the forces between pairs of particles are linearly additive, the above equation applies
between objects of any mass.

In addition, the number of wave modes can be calculated exactly from equation~\ref{eq:28M}
for $G$. Solving for $N$, given $\sigma_m = 1.28 \times 10^{-11}$ g/cm and the known constants,
  \begin{equation}
    N = \frac{c^2}{2 \pi G \sigma_m} = 1.68 \times 10^{38}.
  \end{equation}
$N$ is an important constant in cosmology.

\subsection{Circularly Propagating Wave Modes}
\label{sec-cpwm}

However, it is still difficult intuitively to make sense of the concept of circular propagation
of the modes in an ordinary 3-dimensional space. Fortunately, I was able to prove
mathematically that the wave modes propagate circularly by determining that eigenvectors
corresponding to circularly propagating wave modes exist in a spherical wave system. The
proof depends on the general properties of spherical wave systems and, specifically, on
the assumed $1/r$ dependence of the intensity. The wave modes are actually the normal modes
of the universe and theoretically completely define the universe. They are a new concept
in cosmology and, in fact, their physical existence, to my knowledge, has not previously
even been conjectured.

The proof is as follows: Begin with the classical wave equation in spherical coordinates and
then consider the separated radial wave equation, $R(r)$, given by
\begin{equation}
    \frac{d^2R(r)}{dr^2} + \frac{2}{r} \frac{dR(r)}{dr} +
        \left(k^2 - \frac{l(l + 1)}{r^2}\right)R(r) = 0
    \label{eq:75}
\end{equation}
where $k^2 = (2 \pi f)^2/c^2$.

The standard method~\cite{ST:CC} of solving for $R(r)$ is to separate
equation~\ref{eq:75} into two parts
\begin{equation}
  R(r) = r^{-1/2} B(r)
  \label{eq:rad}
\end{equation}
where $B(r)$ is Bessel's equation of half-integral order. The solutions
of $B(r)$ are non-periodic except when the angular momentum eigenvector
$l = 0$. Since periodicity is essential for constructive interference,
this is the only solution consistent with the wave system.  For this
unique solution, the solution is
  \begin{equation}
    B(r) = \sqrt{\frac{2}{(\pi r)}} \left(\sin{(r)} + \cos{(r)}\right).
    \label{eq:bess}
  \end{equation}
However, since $\cos{(r)}$ equals $1$ at $r = 0$, the factor $\sqrt{(2/(\pi r)}\cos{(r)}$
goes to infinity at $r = 0$. Consequently, only the $\sin{(r)}$ part of the solution can be
used.
However, this is still not the required solution to the problem. Since the intensity $I$ is
proportional to $R^2(r)$, $I$ is proportional to $1/r^2$. As previously discussed, this leads
to an experimentally incorrect $1/r^3$ law for gravitation based on the general force law.

Nevertheless, the problem can be completely solved by noting that the above formulation of the
problem assumes the tension and mass density are uniform for each wave mode. But, this is not true in the wave system since the tension and mass density parameters are assumed
proportional to the local intensity of a wave mode. Since the local
intensity at $r$ for a circularly propagating wave mode is proportional
to $1/r$, we must have $T = T_o/r$ and $\sigma = \sigma_o /r$.

For these parameter variations in the spherical wave equation, the amplitude $B'(r)$
becomes~\cite{GR:PM}
  \begin{equation}
    B'(r) = \frac{1}{(T \sigma)^{1/4}} B(r) = \frac{s^{1/2}}{(T_o \sigma_o)^{1/4}} B(r)
    \label{eq:bess1}
  \end{equation}
since $(T \sigma)^{1/4} = r^{-1/2}(T_o \sigma_o)^{1/4}$. Then the
radial amplitude function becomes
  \begin{equation}
    R(r) = r^{-1/2} B'(r) = B(r)
    \label{eq:R1}
  \end{equation}
and the intensity of $R(r)$ is then proportional to $1/r$, as required. The modes represented
by $R(r)$ in equation~\ref{eq:R1} are very important physically since they are the normal
modes of the universe.

What then are the characteristics of this wave model of the universe?
First, the normal modes of vibration of the universe are confined to
planes orientated in different directions. Each plane is excited as a
single mode by a particle and the resulting wave mode propagates
circularly in a plane centered on the particle. All the particles in
the universe are located on the surfaces of these planes. The three
dimensional character of universe is thus made up of a collection of
$N$ vibrating planes orientated at different angles and these simulate
the isotropic propagation of gravitation.

\subsection{Absorption of Gravitation}

Consider the case of x-rays incident on a perfect crystal.
Experimentally, the x-ray energy density inside the crystal decreases
exponentially with distance due to interactions of the x-ray photons
with particles within the crystal. By analogy, it is proposed that the
intensity of the wave modes are reduced exponentially by absorption and
re-radiation by the mass particles. Since the absorbed energy at each
interaction of a wave mode is re-radiated in many directions other than
the original direction of the wave mode, the intensity of the wave mode
along its original direction is reduced. The gravitational force is
then reduced in proportion to the decrease in intensity.

Using the theory developed in this paper, the absorption of the wave modes originating
from a single proton can be calculated. The energy density of a single mode is given by
\begin{equation}
  \sigma_e  = \frac{mc^2}{2 \pi N s} \exp{(-s/R)}
            =  \frac{1.42 \times 10^{-42}}{s} \exp{(-s/R)} \mbox{~erg/cm}
\label{eq:se}
\end{equation}
where $R = 1.85 \times 10^{28}$ and $N = 1.68 \times 10^{38}$. Note: The variable
$s$ will be used from now on for the euclidean distance and $r$ for the normalized distance,
$s/R$.

Equation~\ref{eq:se} assumes that the energy density decreases as $1/s$
and is additionally reduced by a factor $\exp{(-s/R)}$ due to
absorption by mass particles. Then, the energy, $E$, absorbed by
another proton $s$ distant from the first is
  \begin{equation}
    E  =  \lambda \sigma_e =  \frac{1.85 \times 10^{-55}}{s} \exp{(-s/R)} \mbox{~erg}
  \label{eq:37M}
  \end{equation}
where $\lambda = 1.3 \times 10^{-13}$ cm, the linear size of a proton. $k_o = 1.85 \times
10^{-55}$ is the gravitational absorption constant.

The above absorption theory can be easily applied to a spherical body, such as a star.
Consider a spherical distribution of particles with a specific gravity $d_{sg}$ ($d_{sg} = 1$
for water with a density $1.0$ g/cm$^3$). The total energy absorbed from a single proton at
the center of the sun is the product of $E$ (equation~\ref{eq:37M}) times the number of
protons, $n$, within a radius $s$. $p$ is the number of protons per gram, $5.99 \times 10^{23}$. Then, the total energy absorption is
 \begin{eqnarray}
    E_{Abs} & = & 4 \pi p d_{sg} \int_{0}^{r}\frac{k_o}{s}s^2 dr
             =  2\pi p d_{sg} k_o s^2 \nonumber \\
           & = & 7.0 \times 10^{-31} d_{sg} s^2 \nonumber.
 \end{eqnarray}
Since the initial source energy, $E_i$, is due to a single proton with energy $mc^2$, the
proportion, $P$, of the gravitational energy shielded is
 \begin{eqnarray}
   P & = & \frac{E_{Abs}}{E_i} = \frac{7.0 \times 10^{-31} d_{sg} s^2}{1.50 \times
           10^{-3}}\\
     & = & 4.6 \times 10^{-28} d_{sg} s^2. \nonumber
 \end{eqnarray}
For the sun with a radius of $7 \times 10^{10}$ cm and an average density about 10 times
the specific gravity of water, the proportion of the gravitational energy absorbed is
about $2.3 \times 10^{-5}$. This absorption is negligible and thus confirms current
practice of ignoring the possibility of any gravitational absorption for most astronomical
bodies. However, for the very largest stars with radii $10^{13}$ cm ($10^8$ km), the
gravitational absorption approaches 100\%.

\subsection{Gravity Measurements During a Solar Eclipse}

Since the absorption of gravitation is central to the cosmological theories proposed
in this paper, the only known observational evidence of gravitational absorption is
discussed next. Note that gravitational absorption is not theoretically predicted in
either Newtonian gravitation or in general relativity.

From the gravitational theory proposed in this paper, a small absorption of the sun's
gravitation by the moon is expected. Recently, the vertical acceleration at the earth's surface
during a total solar eclipse in China on March 9, 1997 was accurately measured by
Wang~\cite{W:abs}. The measurements were corrected for the tidal effects of the sun and moon
and for the earth's rotation. These corrections are accurate to a precision
of $1.0 \times 10^{-6}$ cm/sec$^2$. Since the results of the gravitational measurements,
as described below, were quite different than expected, Wang could offer no explanation
of the results but still believed they were a gravitational effect, possibly indicating gravitational shielding by the moon.  However, Unnikrishnan~\cite{GR:CU} has strongly disputed the possibility of shielding of the sun's gravity by the moon based on laboratory gravitational experiments as well as astronomical and planetary observations.

At the time of total eclipse, Wang expected that the measured vertical
acceleration towards the earth would increase since the acceleration
towards the sun would decrease due to the gravitational absorption of
the moon. Contrary to this expectation, the acceleration at the time of
total eclipse only shows a small decrease of less than $1.0 \times
10^{-6}$ cm/sec$^2$. Instead, two much larger, nearly symmetrical
decreases in the measured acceleration occurred before first contact
and after fourth contact.

\begin{figure}
\includegraphics[width=\mypicsize]{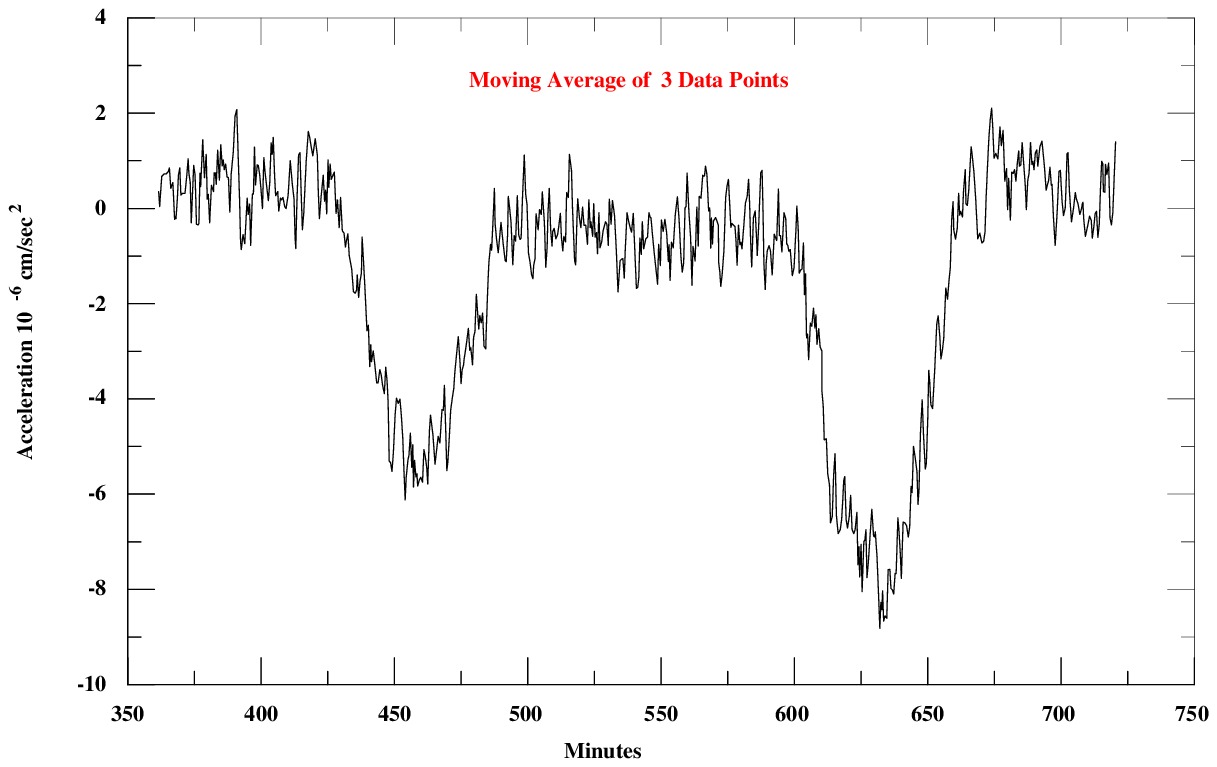}
\caption{Measured vertical gravity variations in China during solar
eclipse of March 9, 1997. Local time shown in minutes from midnight.}
\label{f:fig1}
\end{figure}

The symmetrical decreases in acceleration are clearly shown in Figure~\ref{f:fig1}\ 
which shows the corrected measurements based on a moving average over
three data points.  The first, with a maximum decrease of $5.8 \times
10^{-6}$ cm/sec$^2$, begins about $120$ minutes before totality and
ends about 50 minutes before totality. The second, with a maximum
decrease of $8.7 \times 10^{-6}$ cm/sec$^2$, begins about $50$ minutes
after totality and ends about $120$ minutes after totality. Most
important, these are the only statistically significant changes in the
gravitational acceleration measured over a period of $1/2$ week before
and after the eclipse.

It is believed that these observed effects are due to gravitational absorption based on
the derivation of Newton's law of gravitation. Any absorption of gravity during an eclipse
of the sun by the moon would decrease the acceleration of the earth towards the sun. However,
since the earth and the gravity meter are in free fall around the sun, no change in the
measured acceleration should occur during a total eclipse.

But, the situation is different for the symmetrical reductions in acceleration. The sun is, of
course, eclipsed at other locations before and after the total eclipse at the measurement
location. Because the plates of the earth are rigid, it is proposed that the reduction in
acceleration of the local plate at these other locations is transmitted to the measurement
location and this is what is being measured by the gravity meter. Furthermore, since the
strain due to the acceleration is quite small, it is likely that the local plate is freely
falling because the stress opposing the acceleration is small.

The reduction in acceleration during an eclipse of the sun can be
calculated by the same method used previously to calculate the
absorption of gravitation by the mass of the sun. In this calculation,
assume a bar of matter, representing a cross section of $1$ cm$^2$
through the moon, has an average specific gravity, $\sigma_{sg}$. The
ratio of the absorbed gravitational energy to the incident
gravitational energy per proton in the sun will give the proportion of
the gravitational energy which is absorbed. Then, the reduction in the
gravitational acceleration at the earth during a total eclipse will be
given by this ratio.

Applying this to the solar eclipse, given that the diameter, $l_m$, of the moon is
$3.48 \times 10^8$~cm and assuming an average specific gravity of $4$, we have
\begin{equation}
  E_{Abs} = \frac{p d_{sg} l_m k_o}{s} = \frac{1.48 \times 10^{-22}}{s} \mbox{~erg/cm$^2$.}
\end{equation}
The incident gravitational energy per cm$^2$ on the bar from one proton in the sun is
\begin{equation}
  E_i = \frac{1.5 \times 10^{-3}}{4 \pi s^2} = \frac{1.19 \times 10^{-4}}{s^2} \mbox{~erg/cm$^2$}.
\end{equation}
Then, $P = E_{Abs}/E_i = 1.24 \times 10^{-18}$ s. Since the
gravitational acceleration of the sun at the earth is $0.59$
\mbox{cm/sec$^2$} and the distance to the sun is $1.5 \times 10^{13}$
cm, the reduction in acceleration at the earth during the total eclipse
is given by
\begin{equation}
   a = - 0.59 P = - 11.0 \times 10^{-6} \mbox{~cm/sec$^2$.}
\end{equation}
Since the sun was at most at an angle of about $23^o$, the vertical acceleration
is $a \sin({23)}$ or $- 4.3 \times 10^{-6}$ \mbox{~cm/sec$^2$}.

This result is close to the measured symmetrical reductions in
acceleration. It is, I believe, a direct observational proof that
gravitation is absorbed by matter and justifies using the absorption of
gravitation in cosmology. Also note that on the opposite side of the
earth from the visible eclipse (the dark side), the symmetrical
accelerations should be positive, instead of negative. This provides
another opportunity to test the absorption theory of gravitation
proposed in this paper.

The above explanation also appears to be confirmed by the comparison of the
timing between the optical observations and the events as determined by
the measured vertical acceleration data. Table 1 shows the optical times 
(in universal time) of the eclipseevents~\cite{na:na}, the elapsed minutes 
from 16:00 on March 8, 1997 from optical measurements and from acceleration 
based measurements.

\begin{table}
\begin{center}
Table $1$ \\ \vspace{1mm} Timing of Solar Eclipse Events \\ March
9, 1997  \vspace{1mm}

\begin{tabular}{lcccc} \hline \hline
Eclipse Event    & Optical  & Optical & Accel   & Diff \\
                 &  UT      & Minutes & Minutes & Minutes \\ 
\hline
Start Penumbra   & 23:17:38 & 438    & 430    & -8   \\
1st Contact      & 00:03:29 & 483    & 471    & -12  \\
2nd Contact      & 01:08:18 & 548    &        &      \\
Total Eclipse    &          & 549.5  & 540    & -9 \\
3rd Contact      & 01:11:04 & 551    &        & \\
4th Contact      & 02:19:50 & 620    & 608    & -12  \\
End Penumbra     & 02:58:23 & 658    & 662    &  4\hbox to 0pt{*\hss}  \\
\hline \hline

\end{tabular}

\end{center}
*This difference in time is anomalous because there is still
more penumbra to the east after
the specified ``End Penumbra'' time. The ``End Penumbra'' refers
to the most extreme southern part of the penumbra, not a later
occurring part in the east.
\end{table}

Comparison of the optical times with the acceleration times shows that
the acceleration times are earlier than the optical times. This is
expected since theoretically the acceleration times are $500$ seconds
(\hbox{$8$-$1/3$} minutes) ahead of the optical times. This does not mean that
gravitation propagates instantaneously. Instead, the gravitational
field, like the electrostatic field, has a velocity-dependent
component~\cite{Ca:grav} that cancels the effect of the propagation
delay to first order. The difference in the elapsed times between the
optical times and the times derived from the acceleration data tends to
confirm the theoretical prediction.

\section{Cosmological Effects of the Wave System}

Assuming the wave system is the basic phenomena of the universe, the dominant effects in the
universe must be due to the wave system. These dominant effects are the existence of
elementary particles, the energy of mass particles and photons, the mass density of the
universe, the Hubble redshift and the size and nature of the universe.

\subsection{Mass-Energy of Elementary Particles}

To compute the energy absorption by all the protons in the universe, consider a spherical distribution of protons with a number density $\sigma_n$. It is assumed that the wave modes originating from a single proton interact with all the protons in the universe. Then, the total energy, $E_p$, absorbed from a single proton is given by
\begin{eqnarray}
    E_p    & = & 4 \pi \sigma_n \int_{0}^{\infty}\frac{k_o}{s}\exp{(-s/R)}~s^2 ds
             =  4 \pi k_o \sigma_n R^2 \\
           & = & 3.0 \times 10^{-55}~\sigma_n R^2 \mbox{~erg} \nonumber
 \label{eq:ET}
\end{eqnarray}
where $k_o = 1.85 \times 10^{-55}$ from equation~\ref{eq:37M} and
$\sigma_n$ is the number density of particles.

This result identifies the source of the mass energy of elementary particles with the
mutual interactions between all the particles in the universe. This explanation is
consistent with an equilibrium state of the universe because the input energy to a
particle equals the output energy of a particle.

\subsection{Predicted Mass Density of the Universe}

Given the energy of a proton, $\sigma_n$ can be calculated
if $R$ is known. For $R = c/H$ (section~\ref{sec:HRP}) and $H = 50$ km/sec/Mpc, $R =
1.85 \times 10^{28}$ \mbox{cm}. $R$ is the mean absorption distance in the universe. Then,
$\sigma_n$ is given by
\begin{equation}
   \sigma_n  =  \frac{E_p}{4 \pi k_o R^2} =  1.8 \times 10^{-6} \mbox{~particles/cm$^{3}$}.
\end{equation}
Then, the average mass density of the universe, $\sigma_d$, with
$m_p = 1.67 \times 10^{-24}$ g is given by
\begin{equation}
   \sigma_d = m_p \sigma_n = 3.0 \times 10^{-30} \mbox{~g/cm$^3$}.
\label{eq:md}
\end{equation}
This mass density is a factor about ten times greater than is observed. Previous predictions
were even larger in the expanding universe models and, therefore, prompted the idea of
``missing mass.'' With this prediction, the amount of missing mass is reduced but there is
still a very appreciable difference between theory and observation.

\begin{figure}
\includegraphics[width=\mypicsize]{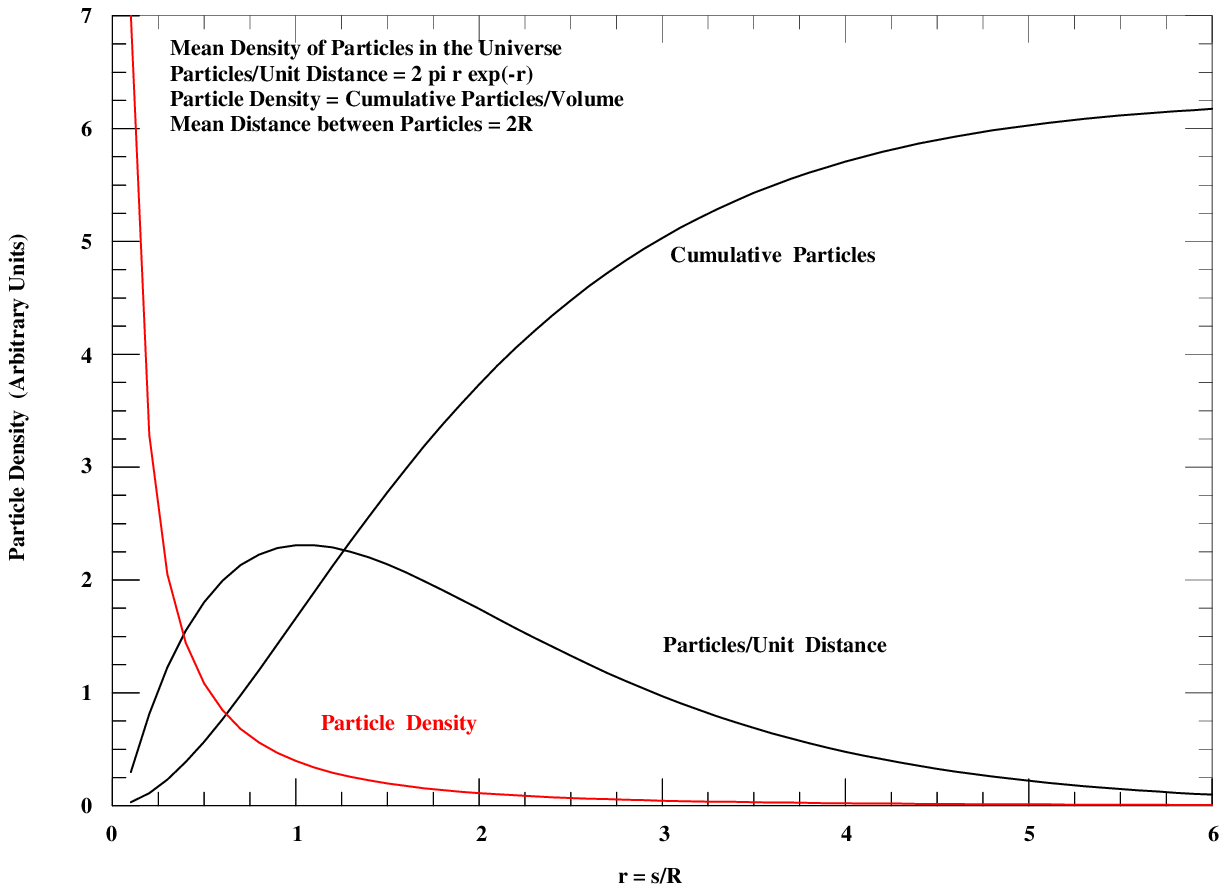}
\caption{Density of particles assuming circularly propagating planes
model of the universe. The mean particle density is given by the
particle density at the mean distance between particles, $s=2R$.}
\label{f:fig2}
\end{figure}

Because of the above inconsistency between theory and observation, it is believed that the
above theory is incorrect. Thus, an alternate theory is proposed.

The structure of the universe is really determined by the circularly propagating planes
which assumes the particles exist only on the surface of the planes. This is a very different
concept from the concept of a random location of particles in a three dimensional space.
The mass density given by equation~\ref{eq:md} was based on this concept and, therefore,
the result is clearly suspect.

Using the concept of circularly propagating planes, the energy of a particle is instead given
by
\begin{equation}
   E_p = 2 \pi N \sigma_s \int_{0}^{\infty}\frac{k}{s}\exp{(-s/R)}s~ds
       =  2 \pi N k_o \sigma_s R. \\
\label{eq:CP}
\end{equation}
From this equation, the surface density of particles, $\sigma_s$, is
\begin{equation}
   \sigma_s = \frac{E_p}{2 \pi N k_o R} = 4.0 \times 10^{-16} \mbox{~particles/cm$^2$}.
\end{equation}
It is difficult to relate this to the previous particle density result of $3.0 \times
10^{-30}$ g/cm$^3$ because one measure is in cm$^3$ and the other in
cm$^2$.

To determine the mean number density (and the mean mass density), first calculate the cumulative number of protons, $n_p$, versus $s$. $n_p$ is given by
\begin{equation}
  n = 2 \pi N \sigma_s \int_{0}^{s} \exp{(-s/R)} s~ds
    = 2 \pi N \sigma_s R^2 \left[1 - (s/R + 1)\exp{(-s/R)}\right].
\label{eq:cm}
\end{equation}
Then, the proton number density can be computed by dividing $n$ by the
Euclidean volume at $s$. This is shown graphically in Figure~\ref{f:fig2}\  (for $N$
and $\sigma_s = 1$). For small $s$, the proton number density is large
but falls rapidly as $s$ increases. The question is "What volume should
be used as a measure of the proton number density?" Since the mean
distance between protons is $s = 2R$, the volume at this mean distance
seems appropriate as a measure of the mean proton number density.
Setting $s$ equal to $2R$, the mean number density of protons is
\begin{equation}
   \sigma_n = \frac{n}{4 \pi/3 (2R)^3} = \frac{3 N \sigma_s}{16R}\left(1 - 3\exp{(-2)}\right)
            = 3.9 \times 10^{-7} \mbox{~particles/cm$^3$}.
\end{equation}
From this result, the mean mass density of the universe is
\begin{equation}
  \sigma_d = 1.67 \times 10^{-24} \sigma_n = 6.6 \times 10^{-31} \mbox{~g/cm$^3$}.
\end{equation}
This mean mass density is close to the observed mean mass density of matter in the local
universe.

\subsection{Size and Nature of the Universe}
\label{sec:SN}

Given a static and equilibrium (non-evolving) universe, the determination of the size and
age of the universe may be logically deduced from the wave system model. A starting point is
the observation that the physical constants~\cite{GR:VV} are exactly the same in very
distant galaxies. Then, assuming the physical constants are a function of the frequency and
number of wave modes, particles in distant galaxies must interact with the same effective
number of distant particles as local particles. However, these distant particles interact
with a different set of particles than the local particles do since all interactions decrease
exponentially with distance due to absorption of the wave modes. Since distant galaxies
interact with even more distant particles than local galaxies, I conclude that the universe
must be very much larger than $R$, the absorption (or interactive) radius of the universe,
or even infinite. Finally, an infinite universe implies an infinitely old universe since
interactions at the velocity of light will take an infinite time to traverse an infinite
universe.

If the static universe is infinite (or very much larger than $R$), some consequences are:
\begin{enumerate}
  \item The gravitational potential must be finite and the same at all points (except near
  large
   masses) since each mass particle interacts gravitationally with the same finite effective
   number of particles. The universe on a large scale then must be flat (Euclidean).
  \item In an infinitely old universe, the universe must be non-evolving on a large scale.
   Therefore, the properties of clusters and galaxies must on the average be independent
   of time and independent of their distance from us. This is another argument
   for the PCP and an equilibrium universe.
  \item A corollary is that processes must exist which maintain the universe in an equilibrium
   state. The determination of these processes should be a goal of future astrophysical
   research.
\end{enumerate}
One process proposed by Moore~\cite{ST:MOOR} could maintain a
non-Maxwell-Boltzman distribution of the particles in an infinite
universe and prevent ``the heat death of the universe''. The process is
based on the principle that the momentum of a gas filling an infinite
volume must be zero with respect to all inertial frames. As a result of
this principle, the infinite system can not reach a Maxwell-Boltzman
equilibrium. Nevertheless, localities do tend towards a localized
equilibrium but this equilibrium is upset at random times by collisions
with other localities. If this process does exist, observers would
always perceive a non-equilibrium state when they look out in space
with telescopes. Then, the ``stable state'' of the universe must be a
dynamic state involving constant change.

\section{Hubble Redshift Process}
\label{sec:HRP}

I assume the Hubble redshift is due to a physical process instead of the expansion of the
universe. However, the derivation of a physical process for the Hubble redshift is a difficult
problem. To my knowledge, all previous attempts to derive a physical process have not been
viable. It is also clear that because the redshift is a global process, the redshift is
telling us something new about the basic workings of the universe. After all, a photon's
energy is reduced by about $2/3$ if it originates at $z = 1$. This reduction in energy,
if applied to an ultraviolet photon, amounts to a large amount of energy. Where does the
energy go? The difficulty of the problem is due to the few clues to the physical nature
of the Hubble redshift.

Only one of the many theorists who attempted to derive a physical process for the Hubble
redshift will be specifically mentioned. Furth~\cite{ST:FUR} hypothesized that gravitational
effects are responsible for the redshift and attempted to link the Hubble redshift to similar
redshifts occurring in the spectral lines near the sun's limb. Specifically, he proposed that
photons forced to move along a curved path in a gravitational field would loose energy. From
this argument, he could show that a redshift approximately proportional to the observed
redshift would occur. However, the complexity of the hypothesis and the uncertainties of
the physical
explanation made the physical process unlikely. Nothing ever came of his hypothesis.

Early on, I concluded that new physics was required because no solution
to the redshift problem had been found, despite the many theorists who
worked on the problem. The Hubble redshift appears to be a basic
problem on a par with other unsolved basic problems of physics. In the
last section, new physics was developed involving the concept of an
underlying wave system. This lead to the concept that the mass energy
of particles is due to interactions between particles.

The energy exchange, $\delta E$, between two particles is proportional to
\begin{equation}
  \delta E \propto \frac{\exp{(-s/R)}}{s}
\label{eq:EXP}
\end{equation}
where $R$ is the mean absorption distance in the universe. The energy of a mass particle or
a photon is, therefore, the sum of these energy exchanges.

Given the above model for the energy of a particle, a local particle moved to another
location will be further from some particles and closer to others. As a result, a net reduction in the energy of the local particle will be shown to occur. It is this reduction in energy
which produces the Hubble redshift of photons.

This model has been simulated on a computer by assuming a particle moves a small
distance $s$ (to the right) along the $x$-axis. The interaction of distant particles
with the local particle is represented by equation~\ref{eq:EXP}. The simulation shows
that the energy of a photon or a mass particle is reduced by
\begin{equation}
  \delta E = - \frac{s}{R}E
\end{equation}
when it moves a distance $s$ in any direction. $E$ is the initial energy of a mass
particle or a photon. The computer program used for the simulation is shown in Appendix A
with a printout of results from one simulation.

\begin{figure}
\includegraphics[width=\mypicsize]{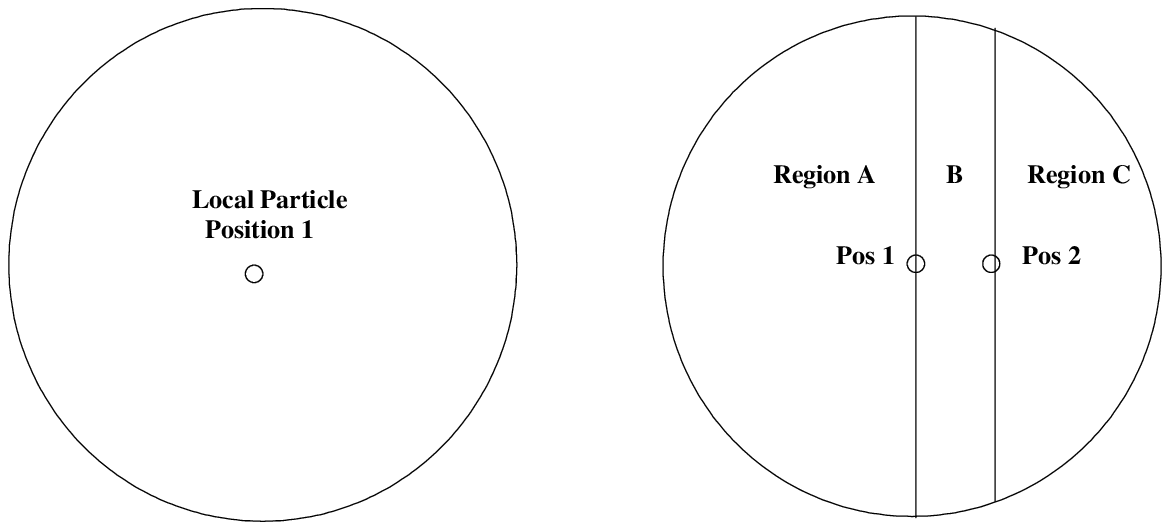}
\caption{Simplified model of the universe. In the left-hand circle, a
local particle in position~1 interacts with all the particles in the
universe. In the right-hand circle, the local particle moves to
position~2 where it interacts with the same particles. The decrease in
the interaction energy is responsible for the Hubble red shift.}
\label{f:fig3}
\end{figure}

However, the simulation is complicated because it takes place in an infinite universe
where the ``cut-off'' of distant particles is soft due to the the exponential term in
equation~\ref{eq:EXP}. Therefore, I have devised a simplified model to understand why
the energy reduction in the simulation occurs.

In the simplified model, the universe is represented by a plane circle with a local
particle at the center. The radius of the circle is large enough $(s > R)$ so that
particles at a greater distance (outside the circle) interact very little with the local particle. The model, therefore, includes essentially all the particles in the universe which significantly interact with the local particle. Furthermore, these interacting particles can be considered fixed in position as the local particle moves. This model is shown in Figure~\ref{f:fig3}A.

For the simulation, the universe is assumed very large and divided into three regions,
A, B and C. This division is shown schematically in Figure~\ref{f:fig3}B. The number of effective particles interacting with the local particle in the simplified model is proportional to the areas of the separate regions in the circle.

Assume the local particle is initially located at position 1 and, subsequently, moves a
distance, $s$ along the $x$-axis to position 2. The reduction in energy when the particle
moves from position 1 to position 2 is determined separately for each region. Wave modes
from two particles located in each region to the local particle will be modeled.
To simplify the bookkeeping, the contribution of energy from each region to the
particle in positions 1 and 2 will be shown in Tables 1A, 1B and 1C.

Mass particles are assumed to result from the constructive interference of wave modes moving in
opposite directions to produce a standing wave. In contrast, a photon is considered a result of
wave modes propagating in the same direction as the photon. Alternatively, the photon can also
be considered a standing wave effect. In any case, both assumptions will be
shown to give the same result for the photon.

First, consider wave modes coming from particles in the left hand region of the circle, labeled
A, and interacting with the particle in position 1. The energy absorbed is proportional to the
area of A. This area is $1/2$ the full area of the universe and contributes energy
$E/2$. This is shown in Table 2A as ``$E/2$'' for region A and position 1.

Wave modes coming from regions B and C also interact with the particle in position 1.
The total energy received from both regions is $E/2$. However, this energy will be separately
accounted for in each region. $E/2(s/R)$ comes from region B and $E/2(1 - s/R)$ comes from
region C. These are shown in Table 2A for position 1 as $E/2(s/R)$ for region B and
$E/2(1 - s/R)$ for region C.

When the particle is in position 2, the particles in region A are at a greater distance from
the local particle. In the simulation, the energy received from each particle in A is given by
\begin{equation}
  \delta E \propto \frac{\exp{\left(-s/R\right)}}{s}
\end{equation}
where $s = \sqrt{(x + s)^2 + y^2 + z^2}$ \mbox{ for $x=>s.$} The energy absorbed from
region A is $E/2(1 - 2s/R)$. The first $s/R$ is due to the greater
distance of the particles. The second $s/R$ is due to the smaller
number of particles since region B is not included in the sum. Finally
the energy received from B is $E/2(s/R)$.

The particles in region C are closer for the particle in position 2. Again, the
energy received from the particles in region C is given by
\begin{equation}
  \delta E \propto \frac{\exp{\left(-s/R\right)}}{s}
\end{equation}
where $s = \sqrt{(x - s)^2 + y^2 + z^2}$ \mbox{ for $x=>s.$} The energy absorbed from region C is $E/2$. This is same as the energy absorbed from region A but this result is due to two compensating effects. The closer particles increase the energy by $E/2(s/R)$ but the fewer particles in region C reduce the energy by the same amount.

\begin{table}
\begin{center}
Table 2A \\

Decrease in Energy of a Mass Particle \\
Moved from Position 1 to Position 2 \\

\vspace{1 mm}

\begin{tabular}{lccc}
\hline \hline
Particle Position  &    Region A    &   Region B  &   Region C         \\
\hline
\vspace{2 mm}
       $P1$        &     $E/2$      & $E/2(s/R)$  &  $E/2(1-s/R)$ \\

       $P2$        &  $E/2(1-2s/R)$ & $E/2(s/R)$  &     $E/2$     \\
\hline
P$2$ less P$1$     &   $-E(s/R)$    &     $0$     &  $E/2(s/R)$  \\
\hline
Net Decrease       &   $-E/2(s/R)$  &             &           \\
\hline  \hline

\end{tabular}

\end{center}
\end{table}

In Table 2A, if the energy associated with the particle in position 2 is subtracted from position 1, the net decrease in energy is $-E(s/R) + E/2(s/R) = -E(s/(2R))$. However, this result must be revised as follows: The input energy to position 2 from region C is nominally $E/2$. But, as in gravitation, the energy absorbed is proportional to the energy of the absorbing particle. The energy of the particle has been reduced already by $E/2(1 - s/R)$, the sum of the energies from regions A and B for position 2, Consequently, the input energy to position 2 from region C is reduced to $E/2(1 - s/R)$. This is shown in Table 2B. Then, the net decrease in energy $(P2 - P1)$ is $-E(s/R)$.

\begin{table}
\begin{center}
Table 2B \\

Revised Decrease in Energy of a Mass Particle \\
Moved from Position 1 to Position 2 \\

\vspace{1 mm}

\begin{tabular}{lccc}
\hline \hline
Particle Position      &      Region A      &   Region B   &     Region C   \\
\hline
\vspace{2 mm}
       $P1$            &      $E/2$         & $E/2(s/R)$   &  $E/2(1-s/R)$ \\

       $P2$            &  $E/2(1-2s/R)$     & $E/2(s/R)$   &  $E/2(1-s/R)$ \\
\hline
P$2$ less P$1$         &    $-E (s/R)$      &     $0$      &       $0$       \\
\hline  \hline

\end{tabular}

\end{center}
\end{table}

For the photon, Table 2C shows the energy changes. As before, the
energy is reduced by $E(2s/R)$ but at the same time there is an
increase in energy from region B. The net decrease is, therefore,
$E(s/R)$. But note that this applies to the total energy of a photon
propagating to the right. Therefore, the net reduction in the energy,
$E$, of a photon is the same as for a mass particle.

\begin{table}
\begin{center}

Table 2C \\

Decrease in Energy of a Photon \\
Moved from Position 1 to Position 2 \\

\vspace{1 mm}

\begin{tabular}{lccc}
\hline \hline
Particle Position      &   Region A   &  Region B  &     Region C      \\
\hline
\vspace{2 mm}
       $P1$            &    $E$       &    $0$     &                   \\

       $P2$            & $E(1-2s/R)$  & $E(s/R)$   &                   \\
\hline
P$2$ less P$1$         &  $-E(2s/R)$  & $E(s/R)$   &                   \\
\hline
Net Decrease           &  $-E(s/R)$   &            &                   \\
\hline  \hline

\end{tabular}

\end{center}
\end{table}

This reduction in energy produces redshifts photons or acts to de-accelerate a mass particle,
assuming the mass particle has kinetic energy. Of course, to move an intially stationary
particle, the ``Hubble'' force (in addition to the ordinary inertial force) is required to
move a mass particle.

\subsection{Photon Redshift}

Let the particle moving to the right be a photon. The reduction in the
energy is given by Table 2C. Consequently, the energy of a photon
decreases as it moves a small distance $ds$. This energy decrease is given by
\begin{equation}
       dE = - \frac{ds}{R}E
\end{equation}
where $E$ is the original energy of the particle and $R = c/H$ is the mean absorption or interaction distance of the wave modes. Integrating the above equation, $\ln{(E/E_o)} = - s/R$ or
$E = E_o \exp{(-r)}$ where the normalized distance $r = s/R$. Then, from the relation,
$f = E/h$, the redshifted frequency, $f$, is given by
\begin{equation}
      f = f_0 \exp{(-r)}.
\end{equation}
For small values of the redshift, $r$ is equal to $z$. For large values of the redshift,
$r = ln(1 + z)$.

This process is proposed as the cause of the observed Hubble redshift of photons. Since the
redshift is a pure gravitational effect, the energy of the photon is reduced without any
blurring of distant galaxies.

\subsection{Mass Particle Redshift}

The redshift has an equal effect on the mass particle. Thus, the decrease in the energy
of the particle (see Table 2B) as it moves a small distance $ds$ is given by
\begin{equation}
  dE = -  \frac{ds}{R}E.
\label{eq:PR}
\end{equation}
This mass particle ``redshift'' possibly explains a part of the observed, very small, anomalous
acceleration toward the sun of the Pioneer 10 and 11 spacecraft~\cite{GR:JA}. An expression for
the acceleration can be derived from equation~\ref{eq:PR} by setting $dE/ds = ma$ and using the
relations $E = mc^2$ and $R = c/H$. Then, the acceleration due to the redshift of mass
particles is
\begin{equation}
  a = - cH.
\end{equation}
This acceleration is directed towards the sun.

For $H = 59$ $(5)$ km/sec/Mpc~\cite{GR:TM}, the predicted acceleration
is $5.6~(0.7) \times 10^{-8}$ cm/sec$^2$. This compares with the recent
result on the anomalous acceleration~\cite{ST:JA} of $8.74$~$(1.25)$
$\times 10^{-8}$ cm/sec$^2$ directed towards the sun for Pioneer $10$
and $11$. The difference between the predicted and observed
accelerations could well be due to a small non-isotropic power
radiation of the Pioner equipment directed away from the sun as
discussed by Scheffer~\cite{ST:SC}.

The mass particle redshift also tends to prevent very large mass accumulations in the universe
since it limits the distance a particle can move, given an initial kinetic energy.

\section{Time-Dilation Process for Supernovae}
\label{sec:TD}

Observations of supernovae by Goldhaber~\cite{GR:GH} show that the period of the light curve
of a supernovae is time-dilated proportional to $(1 + z)$. In the expanding universe model,
this effect is explained as follows: Because photons produced at a later time have to travel a
longer distance to the observer than photons produced at an earlier time, the observed photons
are spread over a longer time interval. Since this process does not occur in a static universe,
it is generally considered that the existence of time-dilation proves that the universe
is expanding.

Normally, if the universe is static, there should be no time-dilation.
This is true for elliptical galaxies since their luminosity is
completely accounted for without consideration of time-dilation. However, the physical
situation is different for supernovae. The luminosity of a supernovae
varies significantly over the period it is visible. Therefore, it is
proposed that the varying luminosity in conjunction with the redshift
produces a time-dilation. This increases the period of the supernovae
in the observer's rest frame and reduces the luminosity of the
supernovae.

Quite by chance, I came across a letter to the editor by Noerdlinger~\cite{GR:ND} that
discusses the time-dilation of quasars. Noerdlinger was concerned with determining the
diameters of quasars from observed fluctuations in luminosity. The following quote from the letter describes his theoretical reasoning:
\begin{quote}
  Think of the fluctuation of any one spectral line as amplitude modulation of a
  carrier. Since waveforms are preserved by the redshift, the maxima in
  amplitude must have the same relation to the oscillations of the
  carrier before and after redshifting, and so must become similarly
  spread out in time. This argument is independent of the cause of the
  redshift. Suppose, now, that an observer near the quasi-stellar object
  sees it fluctuate with period $T_o$, and so would say that it's
  diameter could not exceed $d = cT_o$. The terrestrial observer sees a
  period redshifted to the value $T = T_o(1 + z)$. Thus $T_o = T/(1 + z)$
  and the correct value of $d$ is $d = cT/(1 + z)$.
\end{quote}
This physical theory of time-dilation for quasars is directly applicable to
supernovae. Thus, the large variation in luminosity of a supernovae increases the period,
$T_o$, of the light curve of supernovae by the factor $(1 + z)$.

I have an equivalent physical argument for the time-dilation. Since all
frequencies are reduced by the Hubble redshift, the light curve
modulation frequency is also reduced. This reduction in frequency
corresponds to an increase in the original period, $T_o$, of the light
curve by $(1 + z)$. Thus, the whole light curve of a supernova is
time-dilated as observed in the observer's rest frame. Since this
reduces the number of photons received per second by the observer, the
light intensity is reduced. It follows that the apparent magnitude is
increased by $2.5 \log{(1 + z)}$.

Furthermore, since the above process is independent of the cause of the red-shift, it must
also occur for supernovae in the expanding universe models. However, this creates a fatal
problem for expanding universe models. The expanding universe already incorporates a
time-dilation effect which affects any luminous object equally. Thus, in the expanding universe
model, the time-dilation due to the variation of the supernovae light curve is a second
time-dilation effect.  Since only a single time-dilation effect was observed by Goldhaber,
the expanding universe model must be considered logically falsified. Of course, no such logical
problem occurs in the static universe model. Only one time-dilation effect occurs for
supernovae in the static universe model and this agrees with the observations.

\section{Cosmic Microwave Background Process}
\label{sec:CMB}

To derive the cosmic microwave background (CMB) process, the key assumption proposed by
Hoyle~\cite{GR:HY} is that iron whiskers occur throughout space with a mass density, $\sigma$,
and have the large absorption coefficient, $k = 3 \times 10^7$, in the microwave region.
Outside the microwave region, the iron whiskers have the much smaller absorption coefficient,
$k = 10^4$. Consequently, the iron whiskers do not affect observations by optical or radio
telescopes. If this assumption is correct, it is only necessary  to show that the source of
the energy in the CMB is redshifted visible and ultraviolet starlight.

In this connection, Burbidge~\cite{ST:GB} has shown that if the mass
density is about $3 \times 10^{-31}$ g/cm$^3$ and the He/H ratio by
mass is $0.244$, then the energy released is $4.4 \times 10^{-13}$
erg/cm$^3$. If this energy is thermalized, the CMB black body
temperature is $T = 2.76^o$ K which is close to the observed $2.73^o$
K.

Given that the static universe is infinite in size, it is shown below that visible and
ultraviolet starlight can be redshifted to microwave frequencies with an increase in intensity.
Using the normalized distance $r = s/R$,
\begin{equation}
  I_T = 4 \pi \int_0^\infty \frac{I_o}{4 \pi r^2} \exp{(-r)}~r^2 dr
      = \int_0^\infty I_o \exp{(-r)}~dr = I_o.
\end{equation}
The integration above is over an infinite universe where $I_o$ is the average intensity of
starlight emitted from one cubic centimeter of space. The exponential factor here is due to
the Hubble redshift. Remarkably, this sum equals the initial intensity (ignoring the
small absorption in the visible region), i.e., $I_T = I_o$ except that the redshift
reduces the average frequency by one-half.

Moreover, since the sum, $I_T$, occurs at all points (particles) in an infinite universe,
$I_T$ can be redshifted again to obtain the same intensity which is further reduced in
frequency by one-half. After about $10$ such redshifts, the frequency of $I_T$ is in
the microwave region where the redshifted starlight can be absorbed by the iron whiskers.
Note: The absorption by the iron whiskers in the microwave region limits further integrations.
At the same time, the redshifted light intensity is increased by a factor of $5$ to $10$ by
the multiple redshifts. This process admittedly appears strange but, nevertheless, I believe it does occur.

Assuming the CMB radiation field is in equilibrium, the intensity loss from the CMB field
due to the Hubble redshift must equal the input intensity from the redshifted visible light.
Let the average normalized distance traveled by a CMB microwave photon between emission and absorption equal $r$. Then, the decrease in the CMB intensity is given by
\begin{equation}
  \Delta B = B(1 - \exp{(- r)})
\end{equation}
where $B$ is the CMB radiation intensity. Given that $\Delta B/B \approx 1/25$, $r = 0.04$.
For $B = 1.2 \times 10^{-2}$ \mbox{erg/cm$^2$/sec}, this requires that $\Delta B = 4.8 \times
10^{-4}$ \mbox{erg/cm$^2$/sec}. This intensity, $\Delta B$, is the same order of magnitude
as the total redshifted starlight. Totani~\cite{ST:TOT} cites current measurements of the
extragalactic background light in the near infrared in the range $20$ to
$30$ nW/m$^2/$sr. This is equivalent to $1.0 \times 10^{-4}$ to $1.5 \times 10^{-4}$
erg/cm$^2/$sec. This intensity is slightly smaller than the $\Delta B$ required to maintain
the equilibrium of the CMB. Still, it appears the CMB is in an equilibrium state consistent
with the intensity of light emitted by the stars and the reduction in the CMB intensity
due to the Hubble redshift.

For a single absorption and emission in the distance $r$, the required
mass density of the iron whiskers is given by $\sigma = 1/(t k c)$
where $t = rR/c$. Thus, for $r = 0.04$, $R = 1.85 \times 10^{28}$~cm
and $k = 3 \times 10^7$, $\sigma \approx
10^{-34}$ \mbox{~g/cm$^3$}. While this mass density of iron whiskers
is large, it is possible if the iron whiskers accumulate in
extragalactic space over very long periods of time (say $50 \times
10^{9}$ years) between recycling. Since Hoyle has estimated that the
production of iron by supernovae explosions over $10 \times 10^{9}$ years
results in an iron density of about $10^{-35}$ \mbox{ g/cm$^3$}, an
iron density of $10^{-34}$ \mbox{ g/cm$^3$} seems reasonable.

The number of absorptions and emissions for $r = 0.04$ is on the order of $25$,
more than sufficient to thermalize and smooth the CMB field. Furthermore, the CMB
intensity should be uniform throughout the universe since the iron whiskers are evenly distributed in extragalactic space due to radiation pressure.

\section{Observational Tests}
\label{sec:OT}

The observational tests in this paper include surface brightness, apparent magnitude,
angular size and galaxy counts. These are the same tests suggested by Sandage~\cite{GR:AS95}.

First-rank elliptical galaxies and supernovae are the objects most
useful in determining the space-time metric of the universe. Both have
absolute magnitudes that are nearly the same and thus they are referred
to as standard candles. First-rank elliptical galaxies are more
luminous than supernovae and can measure the metric at larger
distances. On the other hand, supernovae are better standard candles.

Elliptical galaxies of different luminosities and sizes can also be used to measure the
metric by making use of relations between the luminosity, the metric size and the velocity
dispersion of the galaxies. These relations can be determined at low z and then applied to
elliptical galaxies at greater z. These relations are collectively referred to as the
``Fundamental Plane.'' The advantage to using cluster elliptical galaxies is that they
are many times more numerous than first-rank elliptical galaxies and will have different
evolutionary histories.

The data for the tests was obtained from refereed papers published within the last 10 years
(with several exceptions). The data are quantitative, and most important, accurate enough to
distinguish between different models of the universe.

Of course, the observational data must first be corrected by the K-correction. The
K-correction compensates for the changes in the observed spectrum of a galaxy as a function
of the redshift, $z$. Assuming the filter bandwidth is fixed, the K-correction also
includes a correction factor for the observed decrease in luminosity due to the $(1 + z)$
wavelength stretching in the observer's rest frame.

For the static universe model, the observed slope, its standard deviation and the standard
deviation for the individual galaxy observations are shown on each graph. Descriptions of
the mathematical models and the methods used for analysis of the observational data using
the static universe model are given in the following sections.

For various reasons, there are usually a few outliers in a set of data. Outliers are
defined as values which are more than $2.5$ standard deviations from the theoretically
expected value. Since these outliers have a disproportionate effect on the data average,
I discard these outliers. With good data, the number of outliers generally do not exceed $5$\% of the data. When this is the case, I find that the data set is significantly improved without
any compromise in the integrity of the data.

\section{Surface Brightness}
\label{sec:SB}

The surface brightness test will be discussed first. It is the only observational test
which discriminates between any expanding universe model and the static universe model.
Two surface brightness tests were made, each based on a different data set.

The first data set includes only first-rank elliptical galaxies. This data set was compiled
from the observational studies of different observers. Below $z = 0.1$, the data is from
ground based observations. Above $z = 0.1$, the data is from HST observations. All of the
data is based on similar data reduction procedures. K-corrections were generally made by the
observers and were incorporated in their listed data.

The second data set consisting of cluster elliptical galaxies was compiled by
Kochanek~\cite{ST:KOCH100} also from observational studies of different observers.
However, to insure uniformity in the measurement of the galaxy parameters, Kochanek
re-measured the parameters of each galaxy in the different studies. Kolchanek did not
estimate the K-corrections or the galactic absorptions for the individual clusters.
I determined the K-corrections from Fukugita~\cite{ST:MF} and calculated the galactic
absorptions from the foreground galactic extinctions $E(B - V)$ which were provided by
Kochanek and an $R = 3.1$ extinction curve.

In contrast to the first data set, the set from Kochanek consists of elliptical galaxies of
widely different luminosities and physical sizes. Consequently, before the surface brightness
test can be applied, the galaxies must be normalized to a standard elliptical galaxy using the
fundamental plane method. I arbitrarily chose a standard elliptical galaxy with an
effective radius of $10$ Kpc and a velocity dispersion of $225$ Km/sec.

Both sets of data are also used for Euclidean apparent magnitude tests. In addition, the
first-rank elliptical galaxy set is used for the angular size test of the metric. The
two sets of data are listed in Tables 3A, 3B, 4A and 4B.

Figure~\ref{f:fig5}\  shows frequency distributions of the parameters of the first set of data.
The parameters of the first-rank elliptical galaxies are computed based on the static
universe model cosmology. The frequency distributions show nothing unusual about the
test sample of first-rank elliptical galaxies.

Figure~\ref{f:fig7}\  shows a frequency distribution of the parameters of the second set of data
(less 3 outliers). However, for this set, the parameters of the cluster elliptical galaxies
are derived based on the standard elliptical galaxy. Again, the frequency distributions show
nothing unusual about this test sample.

The surface brightness is given by
\begin{equation}
   SB = m_e + 2.5 \log{(\pi \theta^2)}
\end{equation}
where SB (in mag/$\theta^2$) is the surface brightness, $\theta$ is the effective
(half-light) angular radius and $m_e$ is the apparent magnitude within the effective
angular radius.

In the flat, static universe model, the observed surface brightness
varies with redshift as
\begin{equation}
   SB(z) =  2.5 \log{(1 + z)} + SB(0).
\end{equation}
Thus, when the surface brightness versus $2.5 \log{(1 + z)}$ is plotted, the plot
theoretically is linear with slope $1$.

\begin{figure}
\includegraphics[width=\mypicsize]{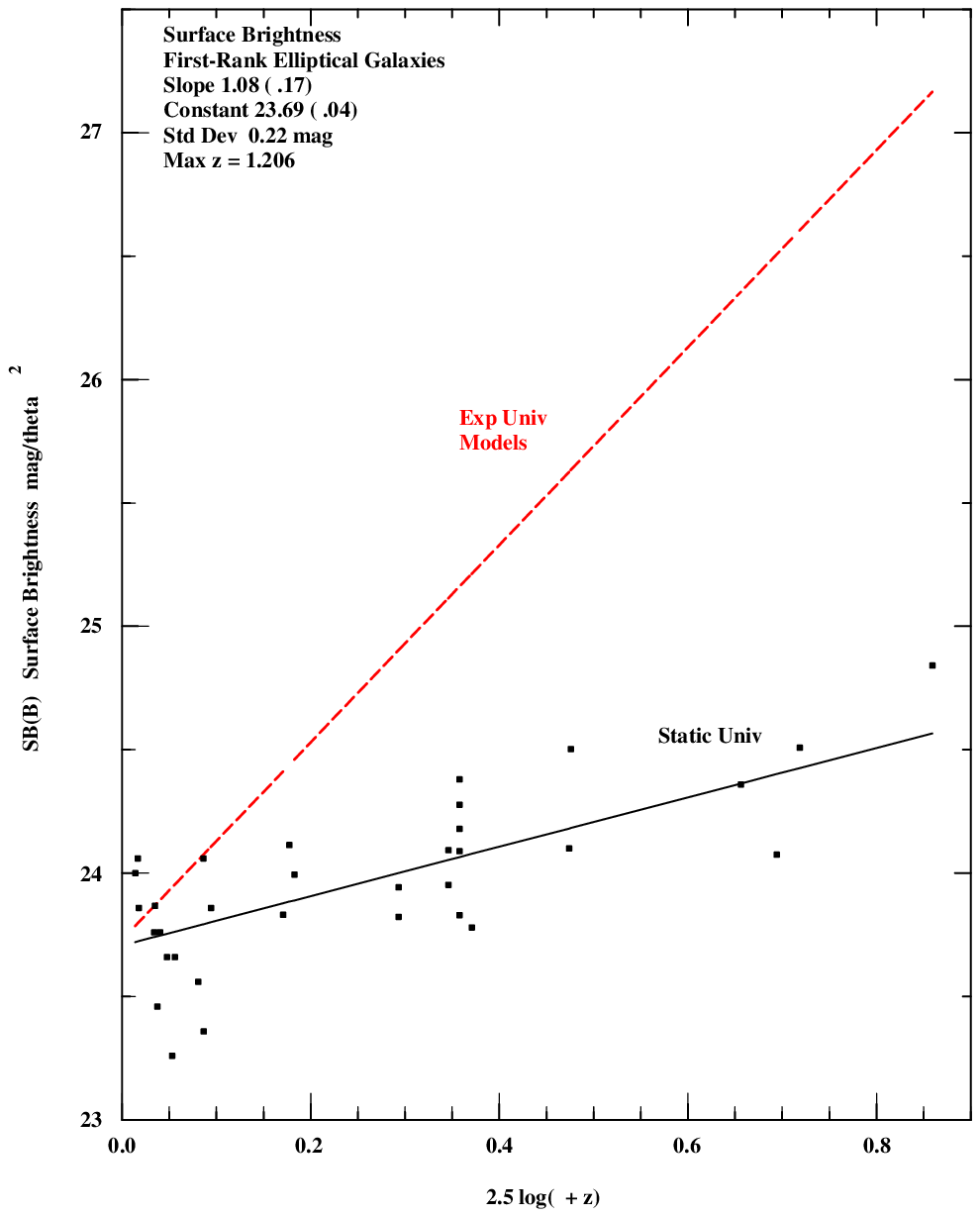}
\caption{Surface brightness of first-rank elliptical galaxies. The
black squares represent the surface brightness observations and the
black line is the static universe regression line with a theoretical
slope of~1. The dashed red line has a theoretical slope of~4 
in expanding universe models.}
\label{f:fig4}
\end{figure}

\begin{figure}
\includegraphics[width=\mypicsize]{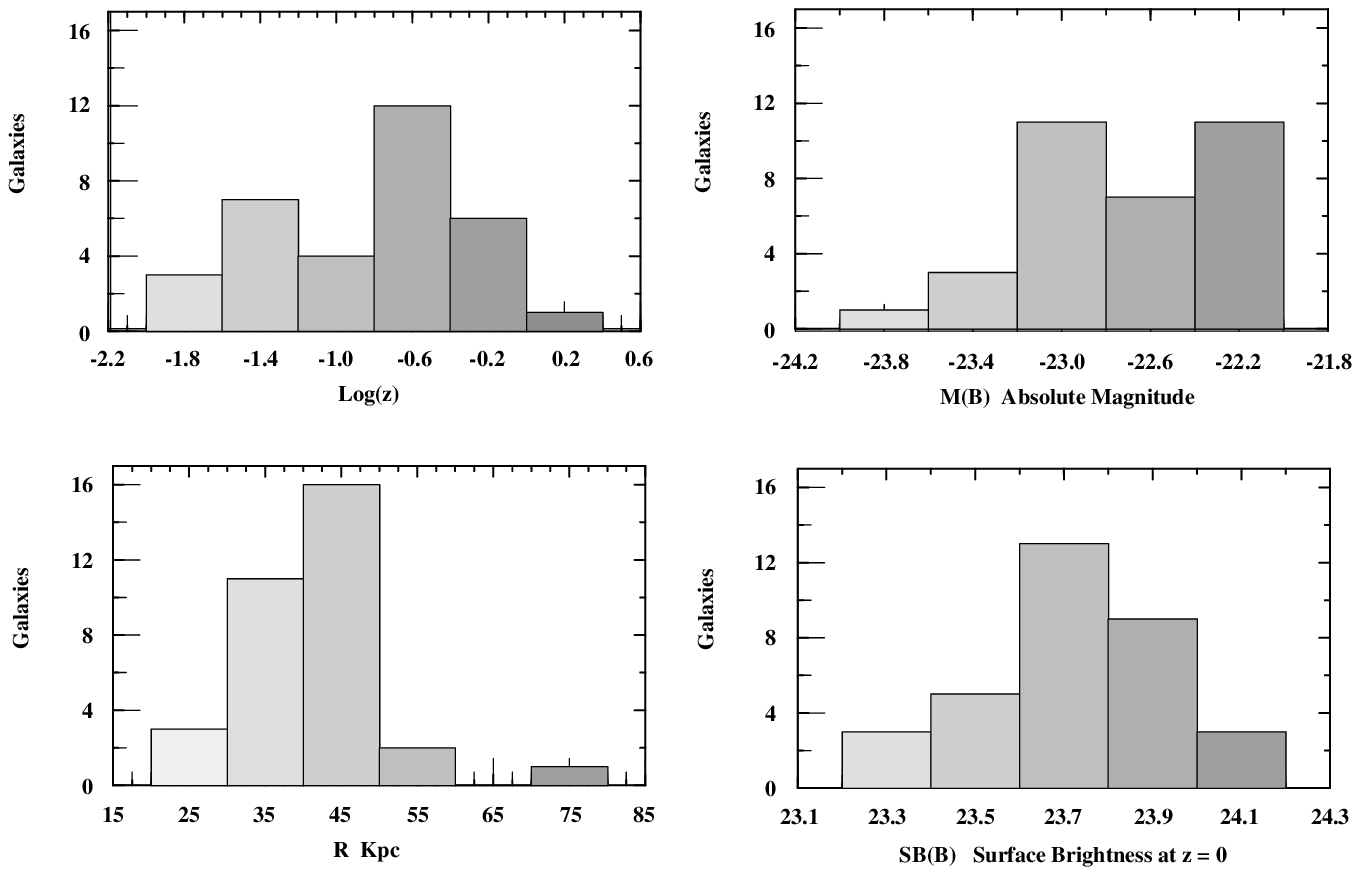}
\caption{Distribution of static universe parameters of test sample of
first-rank elliptical galaxies.}
\label{f:fig5}
\end{figure}

In the expanding universe models, the observed surface brightness varies theoretically as
\begin{equation}
   SB(z) = 4~[2.5 \log{(1 + z)}] + SB(0)
\end{equation}
and, thus, is linear with a slope of $4$. It is important to note that this slope applies
to all of the expanding universe models. Thus, the slope of the surface brightness versus
$2.5 \log(1 + z)$ can clearly determine whether the universe is static or expanding.

\subsection{First-Rank Elliptical Galaxies --- Many Observers}
\label{sec:FRE}

For first-rank elliptical galaxies, the surface brightness test is straightforward because
it is based only on observational data. Since first-rank elliptical galaxies were not
generally identified in the observations, it was assumed the brightest elliptical galaxy
in a cluster was, in fact, a first-rank elliptical galaxy. To further assure that only
first-rank elliptical galaxies were selected, elliptical galaxies were required to have,
in the static universe model cosmology, an absolute magnitude in the Johnson B band brighter
than $-22.2$ magnitudes and an effective physical radius between $25$ and $75$ Kpc. The
restriction on size was to avoid extreme examples of first-rank elliptical galaxies.

The Johnson B band was selected as the common band for the surface
brightness data since most of the galaxies were observed in the B band.
There were some problems encountered in using the data. In many cases,
the expected surface brightness dimming in the expanding universe model
of $10 \log{(1 + z)}$ was subtracted from the observational data and,
therefore, had to be added back to obtain the correct observational
data. In other cases, the surface brightness was calculated from the
listed absolute magnitude and the effective angular radius. And, of
course, data observed in other bands was converted to the Johnson
B-band using the conversion tables of Fukugita~\cite{ST:MF}. After
these conversions were made, the observational data was directly
plotted against $2.5 \log(1 + z)$.

Figure~\ref{f:fig4}\  shows that the surface brightness observations fit the static universe model
very well. The least squares regression for the surface brightness is given by
\begin{equation}
  SB = 1.08~[2.5 \log{(1 + z)}] + 23.69.
\end{equation}
The ordinate in this equation is $2.5 \log{(1 + z)}$. The observed slope is $1.08$ which,
with a standard deviation of $0.21$ magnitudes, quite clearly confirms the static universe
model and excludes the expanding universe models.

\begin{table}
\begin{center}
Table 3A \\ Analysis of First-Rank Elliptical Galaxies\\
\vspace{1mm}
Static Universe Model Parameters
\\
\vspace{1mm}
\footnotesize

\begin{tabular}{llcccccc}
\hline \hline
Galaxy         &      Observer            &    z    &   r   &  SB   &  Data  & Corr & SB(B) \\
               &      \& Ref              &  Data   &       & Data  &  Band  & to B & Corr   \\
\hline
Hydral N3311   & Jorgensen~\cite{ST:IJ95} & 0.0129 & 0.0128 & 23.17 & Gunn g & 0.83 & 24.00 \\
AWM3 N5629     & Sandage~\cite{GR:SG91}   & 0.0152 & 0.0151 & 23.1 & V & 0.96 & 24.06 \\
A262 N708      & Sandage~\cite{GR:SG91}   & 0.0164 & 0.0163 & 22.9 & V & 0.96 & 23.86 \\
A496           & Sandage~\cite{GR:SG91}   & 0.0316 & 0.0311 & 22.8 & V & 0.96 & 23.76 \\
A539 d47       & Jorgensen~\cite{ST:IJ95} & 0.0324 & 0.0318 & 23.04 & Gunn g & 0.83 & 23.87 \\
A2052          & Sandage~\cite{GR:SG91}   & 0.0348 & 0.0342 & 22.5 & V & 0.96 & 23.46 \\
A1139 U6057    & Sandage~\cite{GR:SG91}   & 0.0376 & 0.0369 & 22.8 & V & 0.96 & 23.76 \\
A119           & Sandage~\cite{GR:SG91}   & 0.0446 & 0.0436 & 22.7 & V & 0.96 & 23.66 \\
A85            & Sandage~\cite{GR:SG91}   & 0.0499 & 0.0487 & 22.3 & V & 0.96 & 23.26 \\
A978           & Sandage~\cite{GR:SG91}   & 0.0527 & 0.0514 & 22.7 & V & 0.96 & 23.66 \\
A2255          & Sandage~\cite{GR:SG91}   & 0.0769 & 0.0741 & 22.6 & V & 0.96 & 23.56 \\
A2420          & Sandage~\cite{GR:SG91}   & 0.0823 & 0.0791 & 23.1 & V & 0.96 & 24.06 \\
A1126          & Sandage~\cite{GR:SG91}   & 0.0828 & 0.0796 & 22.4 & V & 0.96 & 23.36 \\
A2440          & Sandage~\cite{GR:SG91}   & 0.0904 & 0.0865 & 22.9 & V & 0.96 & 23.86 \\
A2218          & Barger~\cite{ST:BAR98}   & 0.170 & 0.157  & 23.83  & B & 0.00 & 23.83 \\
A2218-L244     & Jorgensen~\cite{ST:IJ99} & 0.177 & 0.163  & 23.16  & V & 0.96 & 24.12 \\
A665-1150      & Jorgensen~\cite{ST:IJ99} & 0.183 & 0.168  & 23.04  & V & 0.96 & 24.00 \\
AC103(1)       & Barger~\cite{ST:BAR98}   & 0.310 & 0.270 & 23.94 &   B  & 0.00 & 23.94 \\
AC103(2)       & Barger~\cite{ST:BAR98}   & 0.310 & 0.270 & 23.82 &   B  & 0.00 & 23.82 \\
MS1512+36      & Bender~\cite{ST:BEND98}  & 0.375 & 0.318 & 23.95 &   B  & 0.00 & 23.95 \\
A370 20        & Barger~\cite{ST:BAR98}   & 0.375 & 0.318 & 24.09 &   B  & 0.00 & 24.09 \\
CL0024 186     & van Dokkum~\cite{ST:DOKK96} & 0.390 & 0.329 & 23.45 & Gunn g & 0.83 & 24.28 \\
CL1447(1)      & Barger~\cite{ST:BAR98}   & 0.390 & 0.329 & 24.18 &   B  & 0.00 & 24.18 \\
CL1447(2)      & Barger~\cite{ST:BAR98}   & 0.390 & 0.329 & 24.38 &   B  & 0.00 & 24.38 \\
CL1447(3)      & Jorgensen~\cite{ST:IJ99} & 0.390 & 0.329 & 24.09 &   B  & 0.00 & 24.09 \\
CL1447(4)      & Barger~\cite{ST:BAR98}   & 0.390 & 0.329 & 23.83 &   B  & 0.00 & 23.83 \\
Abell 851      & Dickenson~\cite{ST:DICK96}& 0.407 & 0.341 & 23.78 &  B  & 0.00 & 23.78 \\
CL0016         & Schade~\cite{ST:SCH97}    & 0.547 & 0.436 & 24.10 &   B  & 0.00 & 24.10 \\
CL1601(2)      & Barger~\cite{ST:BAR98}    & 0.550 & 0.438 & 24.50 &   B  & 0.00 & 24.50 \\
MS1054 1484    & van Dokkum~\cite{ST:DOKK98}& 0.830 & 0.604 & 23.77 & F814W & 0.59 & 24.36 \\
CL1603-431     & Dickenson~\cite{ST:DICK96}& 0.895 & 0.639 & 24.08 &   B  & 0.00 & 24.08 \\
03.1077        & Schade~\cite{ST:SCH99}    & 0.938 & 0.662 & 24.51 &   B  & 0.00 & 24.51 \\
3C324          & Schade~\cite{ST:SCH97}    & 1.206 & 0.791 & 24.84 &   B  & 0.00 & 24.84 \\
53W002*        & Pascarelle~\cite{ST:PASC}  & 2.390 & 1.221 &       &      &      &       \\
\hline
\end{tabular}
\normalsize

\end{center}
* Angular radius only, surface brightness data unreliable.
\end{table}


\begin{table}
\begin{center}
Table 3B \\
Analysis of First-Rank Elliptical Galaxies\\
\vspace{1mm}
Static Universe Model Parameters\\
\vspace{1mm}
\footnotesize

\begin{tabular}{lccccccccc}
\hline \hline

Galaxy         & Disp & Disp & $\theta_e$ & $R_e$ &  M(B)  &  m$\ast$ & Const & M(B)& m$\ast$\\
               & Data & Stat &    Data    &  Stat &  Stat  & Stat     &  FP   &  FP & FP \\
\hline
Hydral N3311   &      &      &   131.8    & 49.13 & -23.04 & 11.39 & -17.40 & -22.80 & 11.63\\
AWM3 N5629     &      &      &   83.2     & 36.51 & -22.34 & 12.44 & -17.14 & -22.34 & 12.25\\
A262 N708      &      &      &   74.10    & 35.06 & -22.51 & 12.44 & -17.36 & -22.41 & 12.19\\
A496           &      &      &   51.30    & 46.43 & -23.13 & 13.23 & -1.58  & -23.13 & 13.38\\
A539 d47       &      &      &   45.71    & 42.34 & -22.86 & 13.54 & -17.45 & -22.86 & 13.56\\
A2052          &      &      &   40.00    & 39.80 & -23.14 & 13.42 & -17.81 & -23.14 & 13.36\\
A1139 U6057    &      &      &   30.90    & 33.18 & -22.41 & 14.32 & -17.34 & -22.41 & 13.99\\
A119           &      &      &   32.40    & 41.13 & -23.08 & 14.01 & -17.70 & -23.08 & 13.99\\
A85            &      &      &   33.10    & 46.89 & -23.72 & 13.61 & -18.16 & -23.72 & 13.78\\
A978           &      &      &   30.20    & 45.12 & -23.14 & 14.30 & -17.63 & -23.14 & 14.42\\
A2255          &      &      &   20.40    & 43.97 & -23.36 & 14.88 & -17.89 & -23.36 & 14.96\\
A2420          &      &      &   24.50    & 56.37 & -23.41 & 14.97 & -17.58 & -23.41 & 15.41\\
A1126          &      &      &   18.20    & 42.12 & -23.32 & 15.07 & -17.91 & -23.32 & 15.09\\
A2440          &      &      &   15.80    & 39.78 & -22.81 & 15.77 & -17.48 & -22.81 & 15.70\\
A2218          &      &      &    6.69    & 30.56 & -22.33 & 17.54 & -17.38 & -22.33 & 17.09\\
A2218-L244     & 207  & 209  &    8.51    & 40.39 & -22.66 & 17.29 & -17.31 & -22.66 & 17.25\\
A665-1150      & 294  & 298  &   11.38    & 55.64 & -23.48 & 16.54 & -17.67 & -23.48 & 16.96\\
AC103(1)       &      &      &    5.70    & 44.77 & -23.17 & 17.88 & -17.67 & -23.17 & 17.98\\
AC103(2)       &      &      &    5.73    & 45.01 & -23.30 & 17.75 & -17.80 & -23.30 & 17.86\\
MS1512+36      & 290  & 298  &    4.76    & 44.05 & -23.18 & 18.23 & -17.70 & -23.18 & 18.31\\
A370 20        & 334  & 343  &    7.63    & 70.72 & -24.07 & 17.34 & -17.91 & -24.07 & 18.11\\
CL0024 186     & 382  & 393  &    3.99    & 38.22 & -22.56 & 18.92 & -17.29 & -22.56 & 18.92\\
CL1447(1)      &      &      &    4.26    & 40.76 & -22.79 & 18.69 & -17.43 & -22.79 & 18.65\\
CL1447(2)      &      &      &    3.95    & 37.87 & -22.43 & 19.05 & -17.18 & -22.43 & 18.91\\
CL1447(3)      &      &      &    3.78    & 36.16 & -22.62 & 18.86 & -17.43 & -22.62 & 18.65\\
CL1447(4)      &      &      &    3.07    & 29.40 & -22.43 & 19.05 & -17.54 & -22.43 & 18.54\\
Abell 851      &      &      &    3.00    & 29.80 & -22.53 & 19.03 & -17.62 & -22.53 & 18.55\\
CL0016         &      &      &    3.57    & 45.31 & -23.22 & 18.87 & -17.70 & -23.22 & 18.99\\
CL1601(2)      &      &      &    3.86    & 49.21 & -23.00 & 19.10 & -17.36 & -23.00 & 19.34\\
MS1054 1484    & 330  & 348  &    1.88    & 33.04 & -22.46 & 20.34 & -17.40 & -22.46 & 20.01 \\
CL1603-431     &      &      &    1.54    & 28.54 & -22.46 & 20.46 & -17.61 & -22.46 & 19.91 \\
03.1077        &      &      &    1.74    & 33.49 & -22.40 & 20.59 & -17.32 & -22.40 & 20.28 \\
3C324          &      &      &    1.93    & 44.42 & -22.82 & 20.56 & -17.38 & -22.82 & 20.65 \\
53W002*        &      &      &    1.1     & 39.07 &        &       &        &        &       \\
\hline   \hline
\end{tabular}
\normalsize

\end{center}
* Angular Radius only, surface brightness data unreliable.
\end{table}

Table 3C shows the average values of the surface brightness at zero redshift (SB less
$2.5\log{(1 + z)}$), the effective physical radii and the absolute magnitudes for both
low and high z galaxies. These averages show that the three parameters are nearly independent
of the
redshift. These results confirm that first-rank elliptical galaxies do not evolve with
redshift as predicted by the PCP and as hypothesized for the static universe.

\begin{table}
\begin{center}
Table 3C \\

First-Rank Elliptical Galaxies\\
Average Parameters Versus Redshift\\
\vspace{1 mm}

\footnotesize

\begin{tabular}{lcccc}
\hline \hline
Galaxy Grouping                      & \# Galaxies &  SB(B) &  M(B) & $R_e$ \\
                                     &            & $z = 0$ &  Stat &  Stat \\
\hline
$z = 0.00$ to $0.177$                &  16        &  23.69   & -22.95 & 41.8 \\
Standard Deviation                   &            &   0.24   &   0.40 &  6.2 \\
$z  = 0.183$ to $1.206$              &  17        &  23.72   & -22.88 & 40.91 \\
Standard Deviation                   &            &   0.20   &   0.46 & 10.3 \\
\hline
All Galaxies                         &  33        &  23.71   & -22.91 & 41.6 \\
Standard Deviation                   &            &   0.22   &   0.43 &  8.6 \\
\hline   \hline
\end{tabular}
\normalsize
\end{center}
\end{table}

\subsection{Cluster Elliptical Galaxies --- Kochanek}
\label{sec:CEG}

As previously indicated, the cluster elliptical galaxies vary widely in luminosity and
physical size. In order to do the surface brightness test, the elliptical galaxies must be
referenced to a (arbitrary) standard elliptical galaxy. Fortunately, this can be done using
the fundamental plane method. The fundamental plane is based on the empirical observation
that the surface brightness, the log of the effective physical radius and the log of the
velocity dispersion are closely related by the linear equation
 \begin{equation}
   SB = -3.76 \log{(\sigma)} + 3.03 \log{(R_e)} + \mbox{C}
   \label{eq:SB}
 \end{equation}
where $\sigma$ is the velocity dispersion, $R_e$ is the effective (half-light)
radius and $C$ is a constant.

Kochanek re-analyzed observational data from Jorgensen~\cite{ST:IJ95} on local elliptical
galaxy clusters assuming an expanding universe with $q = 0.5$ and $H = 50$ km/sec/Mpc and
found that $\mbox{C} = 26.25$ for the F606W band. Using Kochanek's value of $C$ and assuming
the static
universe model and the F814W band, I found that $\mbox{C} = 25.42$.

The fundamental plane is applied as follows:

\begin{enumerate}

\item $R_E$ is calculated from the effective angular radius, $\theta$, and the normalized
distance, $r$, assuming $H = 50$ Km/sec/Mpc using the equation
\begin{equation}
  R_e = 29.09~r \theta.
\end{equation}
\item The velocity dispersion, $\sigma$, was measured using an aperture with
a physical diameter equivalent to $3.4$ arcsec projected on to a galaxy
in Coma. Since the physical diameter so defined depends on the
cosmological model at higher $z$, the physical diameters were converted
to equivalent static universe diameters, assuming a $q = 0.5$ expanding
universe model was used initially to calculate the diameters. Then, the
velocity was normalized to the static universe model using the
following equation~\cite{ST:IJ95A}
\begin{equation}
  \log{(\sigma_{static})} = \log{(\sigma_{exp})} + 0.04~\log{\left(\frac{R_{static}}
  {R_{exp}}\right)}.
\end{equation}
As a result, the measured dispersion velocities are larger in the static universe model.
\item Using the static universe parameters of each elliptical galaxy and with the
surface brightness reduced to the rest frame $z = 0$ by subtracting
$2.5 \log{(1 + z)}$, the constant $C$ in equation~\ref{eq:SB} is calculated for each galaxy.
\item Then, assuming a standard elliptical galaxy (with specified values of
   the parameters), the rest frame surface brightness of each galaxy is calculated from the
   previously determined values of the constant, $C$.
\item Finally, the surface brightness of the galaxy in the original redshifted frame
   is calculated by adding $2.5 \log{(1 + z)}$ to the rest frame surface brightness.

\end{enumerate}

\begin{figure}
\includegraphics[width=\mypicsize]{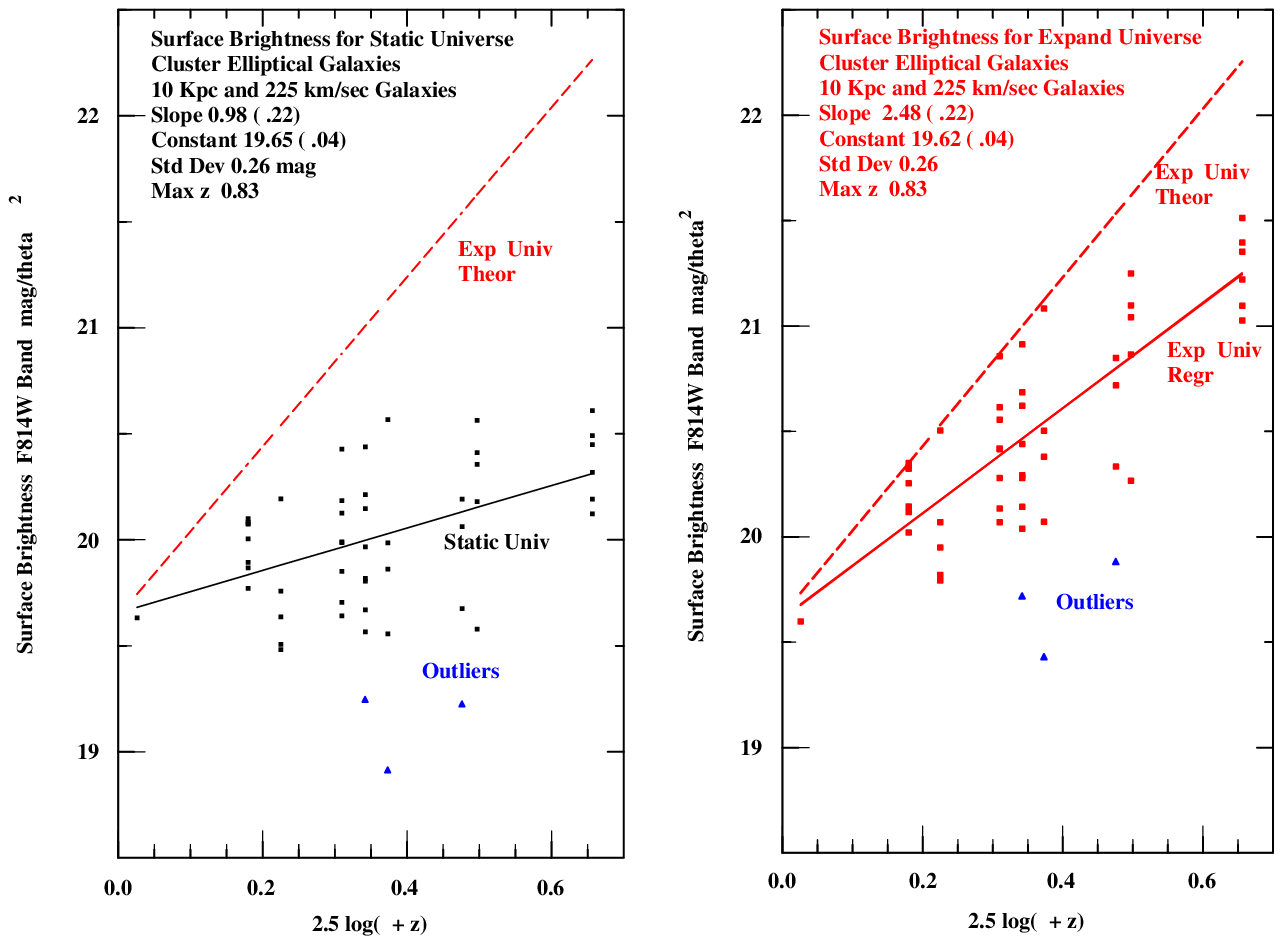}
\caption{Surface brightness of cluster elliptical
galaxies. Fundamental plane analysis method used for both graphs. The
black line with a slope of~1 represents the theoretical static
universe model and the red line is the regression line for the
expanding universe observations. The dashed red line represents the
theoretical expanding universe models.}
\label{f:fig6}
\end{figure}

\begin{figure}
\includegraphics[width=\mypicsize]{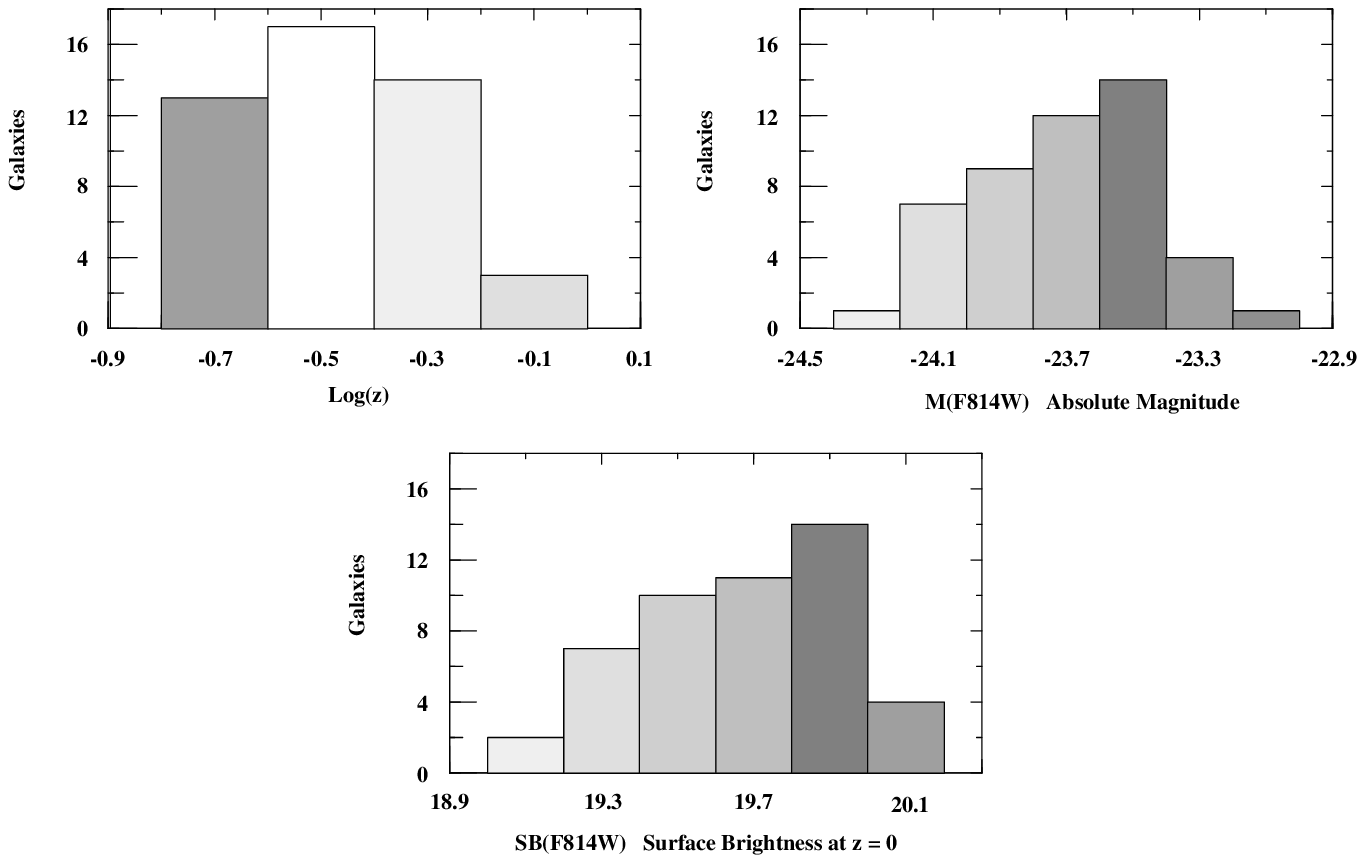}
\caption{Distribution of static universe parameters of test sample of
cluster elliptical galaxies. The parameters were derived from the
data using the fundamental plane method and are based on a standard
elliptical galaxy with a 10~Kpc radius and a velocity dispersion of 225~Km/s.}
\label{f:fig7}
\end{figure}

For the standard elliptical galaxy with $R_e = 10$ Kpc and a velocity dispersion of
$225$ Km/sec, the regression equation for the surface brightness in the F814W band is
\begin{equation}
  SB = 0.98~[2.5 \log{(1+z)}] + 19.65.
\end{equation}
The slope has a standard deviation of only $0.22$ magnitudes. This relation is plotted in
Figure~\ref{f:fig6}\  (left hand panel). It again shows that surface brightness observations fit the static
universe model.

However, this result must be considered suspect since the fundamental plane method is
circular, i.e., the test will tend to support whatever cosmological model is used to
calculate the effective physical radius.

For example, the test shown in Figure~\ref{f:fig6}\  (right hand panel) uses effective physical radii
calculated assuming an expanding universe with $q = 0.5$. The resulting slope of $2.48$
(plus evolutionary brightening) tends to confirm the expanding universe model.

Consequently, the use of only tests based on the fundamental plane method can
not be relied upon to decide between the static universe and expanding universe models.
However, once the type of universe is determined by other tests not susceptible to the
circularity, i.e., the surface brightness test using first-rank elliptical galaxies,
the fundamental plane method can be used as a tool to refine the static universe parameters
of elliptical galaxies.

Table 4C shows the average values of the surface brightness at zero redshift (surface
brightness less $2.5 \log{(1 + z)}$) and the absolute magnitudes for three ranges of $z$.
Note $R_e$ is fixed at $10$ Kpc and the velocity dispersion at $225$ km/sec. These averages
show that the two parameters are nearly independent of the redshift. This confirms that
cluster elliptical galaxies do not evolve with redshift.


\begin{table}
\begin{center}
Table 4A\\
\vspace{1mm}
Fundamental Plane Analysis of Cluster Elliptical Galaxies\\
\vspace{1mm}
Cluster Data and Corrections\\
\vspace{1mm}
\footnotesize

\begin{tabular}{lcccccc}
\hline \hline
Cluster    &   z  &    r  &  Abs  & K-Corr & Filter & Corr to \\
           &      &       &       &        &  Band  &  F814W \\
\hline
Local FP   & 0.024 & 0.024 &      &        & F606W & -0.94 \\
A665       & 0.18  & 0.166 & 0.14 & 0.15   & F814W & 0.00  \\
A2390      & 0.23  & 0.207 & 0.35 & 0.20   & F814W & 0.00  \\
CL 1358+62 & 0.33  & 0.285 & 0.07 & 0.28   & F814W & 0.00  \\
A370       & 0.37  & 0.315 & 0.10 & 0.64   & F675W & -0.59 \\
A370       & 0.37  & 0.315 & 0.10 & 0.32   & F814W & 0.00  \\
A851       & 0.41  & 0.344 & 0.05 & 0.56   & F702W & -0.47  \\
A851       & 0.41  & 0.344 & 0.05 & 0.35   & F814W & 0.00  \\
MS 0015+16 & 0.55  & 0.438 & 0.17 & 0.51   & F814W & 0.00  \\
MS 2053-04 & 0.58  & 0.457 & 0.26 & 0.57   & F814W & 0.00  \\
MS 1054-03 & 0.83  & 0.604 & 0.07 & 1.11   & F814W & 0.00 \\
\hline   \hline
\end{tabular}
\normalsize

\end{center}
\end{table}

\begin{table}
\begin{center}
Table 4B \\
\vspace{1 mm}

Fundamental Plane Analysis of Cluster Elliptical Galaxies\\
\vspace{1mm}
Static Universe Model Parameters\\
\vspace{1 mm}

\footnotesize

\begin{tabular}{lccccccccc}
\hline \hline
Cluster   &  SB   &  SB   & Disp & $\theta$ & $R_e$ & Const &  SB   &   M    & $m^\ast$ \\
Galaxy    & Data  & Corr  & Stat &    Data  &  Stat &  FP   &  FP   &   FP   &  FP      \\
\hline
Local FP  &       &       &      &          &       & 25.42 & 19.63 & -23.96 & 11.81 \\
A665      & & & & & & & & \\
 3        & 19.78 & 19.49 & 277  &   2.19   & 10.53 & 25.40 & 19.76 & -23.98 & 16.00 \\
15        & 19.70 & 19.41 & 262  &   1.51   & 7.29  & 25.71 & 20.07 & -23.67 & 16.31 \\
26        & 19.28 & 18.99 & 228  &   1.07   & 5.16  & 25.52 & 19.89 & -23.86 & 16.13 \\
42        & 19.21 & 18.92 & 250  &   1.05   & 5.04  & 25.63 & 20.00 & -23.75 & 16.24 \\
57        & 18.73 & 18.44 & 213  &   0.66   & 3.18  & 25.49 & 19.86 & -23.89 & 16.10 \\
61        & 18.93 & 18.64 & 230  &   0.71   & 3.41  & 25.73 & 20.09 & -23.65 & 16.33 \\
77        & 20.39 & 20.10 & 150  &   1.29   & 6.20  & 25.70 & 20.07 & -23.68 & 16.31 \\
80        & 19.12 & 18.83 & 190  &   0.66   & 3.18  & 25.70 & 20.06 & -23.68 & 16.31 \\
A2390     & & & & & & & & & \\
6         & 19.67 & 19.12 & 208  &   1.02   & 6.16  & 25.22 & 19.63 & -24.16 & 16.31 \\
7         & 20.16 & 19.61 & 191  &   1.48   & 8.91  & 25.09 & 19.50 & -24.29 & 16.18 \\
9         & 18.63 & 18.08 & 237  &   0.62   & 3.71  & 25.06 & 19.47 & -24.32 & 16.15 \\
10        & 19.33 & 18.78 & 180  &   0.60   & 3.63  & 25.34 & 19.75 & -24.04 & 16.43 \\
138       & 20.87 & 20.32 & 108  &   0.74   & 4.46  & 25.77 & 20.18 & -23.61 & 16.86 \\
CL1358+62 & & & & & & & & & \\
236       & 19.69 & 19.34 & 168  &   0.58   & 4.77  & 25.34 & 19.83 & -24.04 &  17.13 \\
256       & 19.43 & 19.08 & 277  &   0.98   & 8.11  & 25.19 & 19.69 & -24.18 &  16.98 \\
269       & 19.27 & 18.92 & 347  &   0.83   & 6.90  & 25.62 & 20.11 & -23.76 &  17.40 \\
298       & 19.31 & 18.96 & 284  &   0.74   & 6.15  & 25.48 & 19.98 & -23.90 &  17.27 \\
375       & 20.96 & 20.61 & 306  &   2.45   & 20.36 & 25.67 & 20.17 & -23.70 &  17.46 \\
408       & 19.02 & 18.67 & 269  &   0.40   & 3.30  & 25.92 & 20.41 & -23.46 &  17.70 \\
454       & 21.08 & 20.73 & 173  &   1.55   & 12.85 & 25.48 & 19.97 & -23.90 &  17.28 \\
470       & 19.91 & 19.56 & 188  &   0.91   & 7.57  & 25.13 & 19.63 & -24.25 &  16.92 \\
A370      & & & & & & & & & \\
1         & 21.51 & 20.18 & 336  &   2.14   & 19.58 & 25.42 & 19.95 & -23.96 &  17.42 \\
\hline

\end{tabular}
\normalsize
\end{center}
\centerline{Table 4B continued on next page.}
\end{table}

\begin{table}
\begin{center}
Table 4B (Continued)\\
\vspace{1 mm}

\footnotesize

\begin{tabular}{lccccccccc}
\hline \hline
Cluster   &  SB   &  SB   & Disp & $\theta$ & $R_e$ & Const &  SB  & M  & $m^\ast$ \\
Galaxy    & Data  & Corr  & Stat &    Data  &  Stat &  FP   &  FP  & FP &  FP \\
\hline

2         & 23.42 & 22.09 & 258  &   8.91   & 81.62 & 25.02 & 19.55 & -24.36 &  17.02 \\
10        & 21.66 & 20.33 & 199  &   1.41   & 12.94 & 25.26 & 19.79 & -24.12 &  17.26 \\
24        & 20.51 & 19.18 & 255  &   0.79   & 7.27  & 25.27 & 19.80 & -24.11 &  17.28 \\
28        & 19.92 & 19.50 & 225  &   0.68   & 6.19  & 25.60 & 20.15 & -23.78 &  17.60 \\
41        & 19.39 & 18.97 & 298  &   0.51   & 4.70  & 25.90 & 20.42 & -23.48 &  17.90 \\
67        & 21.05 & 20.63 & 164  &   1.02   & 9.37  & 25.67 & 20.20 & -23.71 &  17.67 \\
77        & 21.71 & 20.38 &  96  &   0.91   & 8.35  & 25.70 & 19.23 & -24.68 &  16.70 \\
79        & 20.28 & 18.95 & 171  &   0.46   & 4.19  & 25.12 & 19.65 & -24.26 &  17.13 \\

A851      & & & & & & & & & \\
23        & 20.31 & 19.23 & 191  &   0.65   & 6.45  & 24.98 & 19.54 & -24.40 &  17.17 \\
57        & 19.91 & 18.83 & 203  &   0.41   & 4.07  & 25.29 & 19.84 & -24.09 &  17.48 \\
69        & 20.43 & 20.03 & 197  &   0.58   & 5.75  & 26.99 & 20.55 & -23.39 &  18.18 \\
102       & 19.78 & 18.70 & 152  &   0.23   & 2.34  & 25.41 & 19.97 & -23.97 &  17.60 \\
111       & 21.36 & 20.28 & 59   &   0.55   & 5.49  & 24.34 & 18.90 & -25.04 &  16.53 \\

MS 0015+16 & & & & & & & & & \\
2         & 23.84 & 23.16 & 264  &   10.23  & 130.5 & 25.38 & 20.04 & -24.00 &  18.10 \\
7         & 19.97 & 19.29 & 200  &   0.51   & 6.54  & 24.99 & 19.65 & -24.39 &  17.71 \\
13        & 19.69 & 19.01 & 270  &   0.41   & 5.19  & 25.51 & 20.17 & -23.87 &  18.23 \\
56        & 15.68 & 15.00 & 222  &   0.03   & 0.40  & 25.54 & 19.20 & -24.84 &  17.26 \\
MS 2053-04 & & & & & & & & & \\
197       & 21.59 & 20.76 & 327  &   1.58   & 21.09 & 25.70 & 20.39 & -23.68 &  18.52 \\
311       & 20.45 & 19.62 & 228  &   0.38   & 5.06  & 25.86 & 20.54 & -23.52 &  18.67 \\
422       & 20.05 & 19.22 & 135  &   0.31   & 4.11  & 24.87 & 19.55 & -24.51 &  17.68 \\
432       & 20.96 & 20.13 & 165  &   0.50   & 6.67  & 25.47 & 20.15 & -23.91 &  18.28 \\
551       & 19.56 & 18.73 & 222  &   0.22   & 2.91  & 25.65 & 20.33 & -23.73 &  18.46 \\
MS 1054-03 & & & & & & & & & \\
1294      & 21.35 & 20.17 & 326  &   0.66   & 11.61 & 25.73  & 20.57 & -23.65 & 19.15 \\
1359      & 20.39 & 19.21 & 232  &   0.30   & 5.31  & 25.25  & 20.09 & -24.13 & 18.66 \\
1405      & 21.87 & 20.69 & 267  &   0.95   & 16.79 & 25.44  & 20.28 & -23.94 & 18.86 \\
1457      & 21.42 & 20.24 & 216  &   0.58   & 10.12 & 25.31  & 20.16 & -24.07 & 18.73 \\
1484      & 22.24 & 21.06 & 340  &   1.55   & 27.23 & 25.57  & 20.41 & -23.81 & 18.99 \\
1567      & 21.09 & 19.91 & 269  &   0.47   & 8.22  & 25.61  & 20.46 & -23.77 & 19.03 \\
\hline   \hline

\end{tabular}
\normalsize
\end{center}
*For standard elliptical galaxy: $R_e = 10$ Kpc and $\sigma = 225$ km/sec.
\end{table}

\begin{table}
\begin{center}
Table 4C \\
Cluster Elliptical Galaxies\\
Average Parameters Versus Redshift\\
(10 Kpc and 225 km/sec)\\
\vspace{1 mm}

\footnotesize

\begin{tabular}{lcccc}
\hline \hline
Galaxy Grouping                   & \# Galaxies & SB(F814W) & M(F814W)  \\
                                    &             & $z = 0$  &  Stat  \\
\hline
$z = 0.00$ to $0.23$                &  14        &  19.66   & -23.91  \\
Standard Deviation                  &            &   0.23   &   0.23  \\
$z = 0.33$ to $0.37$                &  15        &  19.61   & -23.96  \\
Standard Deviation                  &            &   0.29   &   0.29  \\
$z = 0.41$ to $0.83$                &  18        &  19.66   & -23.91  \\
Standard Deviation                  &            &   0.28   &   0.28  \\
\hline
All Galaxies (Less 3 outliers)      &  47        &  19.64   & -23.92  \\
Standard Deviation                  &            &   0.27   &   0.27  \\
\hline   \hline
\end{tabular}
\normalsize
\end{center}
\end{table}

%
%
\begin{figure}
\includegraphics[width=\mypicsize]{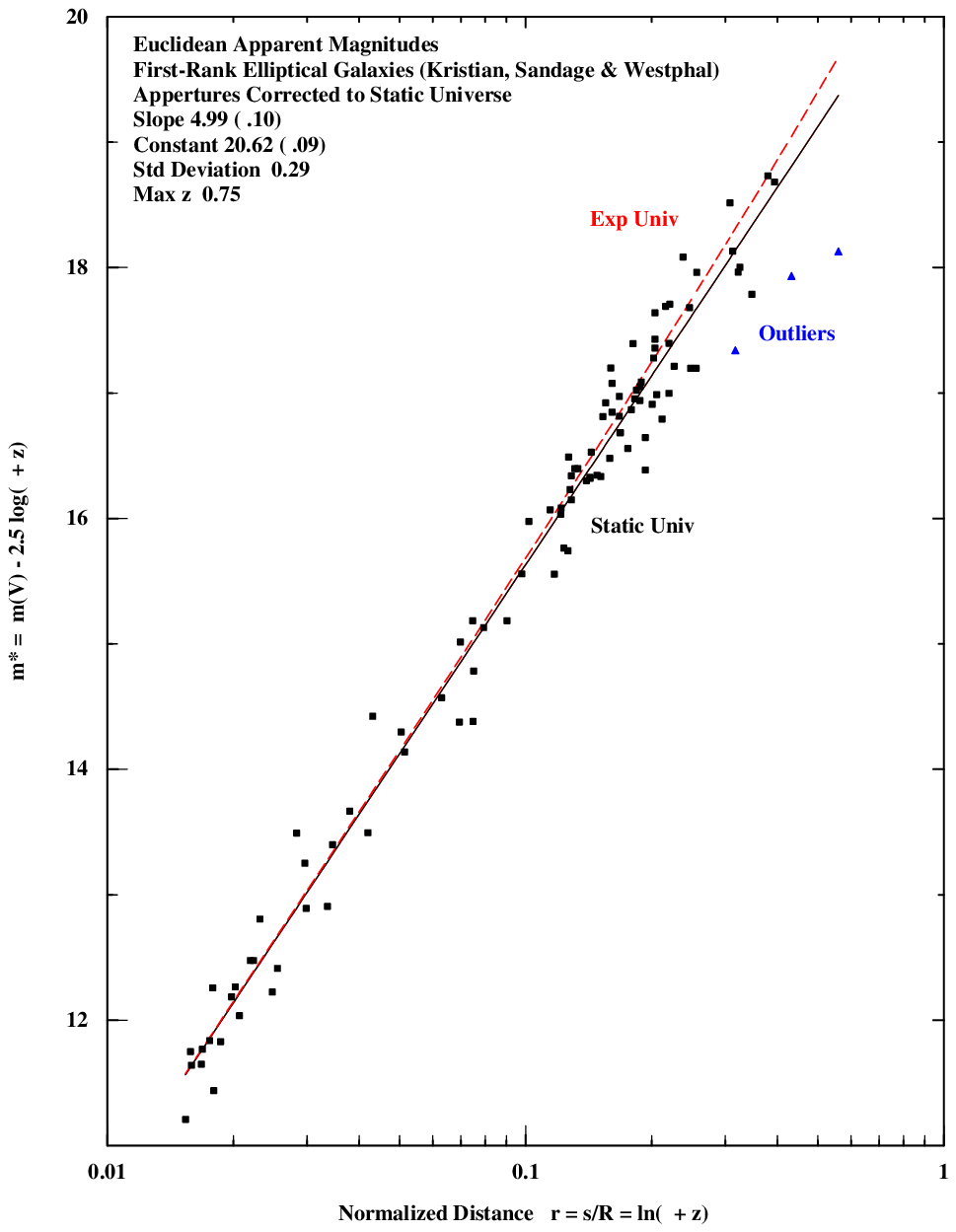}
\caption{Euclidean apparent magnitudes of first-rank elliptical
galaxies. Black line is the static universe regression line with
slope~1. The red dashed line represents the theoretical $q=0.5$
expanding universe model.}
\label{f:fig8}
\end{figure}

\section{Euclidean Apparent Magnitude}
\label{sec:EAM}

The observed luminosity, $l$, of a galaxy in the flat static universe model after
K-corrections and corrections for galactic absorptions is given by
\begin{equation}
   l = \frac{L}{4 \pi s^2 (1 + z)}
   \label{eq:lum}
\end{equation}
where $L$ is the absolute luminosity of the galaxy, $s$ is the Euclidean distance and the
factor $(1 + z)$ accounts for the loss of energy due to the Hubble redshift.

Then, the equation for the apparent magnitude, $m$, for the flat static universe model is
given by
\begin{equation}
    m = M + 5\log{r} + 2.5\log{(1 + z)} + \mbox{C}
    \label{eq:mag}
\end{equation}
where $r$ is the normalized Euclidean distance, $r = s/(2R) = \ln{(1 + z)}$. $R$ is the mean
interactive radius of the universe, equal to $\approx 8.5$ billion light years for a Hubble
constant of $H = 59$ km/sec/Mpc. The absolute magnitude, $M$, is assumed constant since no
evolution occurs in the static universe model.

For the expanding universe models, the apparent magnitude, $m$, is generally plotted versus
$\log{z}$. Instead, the ``Euclidean apparent magnitude'', defined as
\begin{equation}
   m^\ast = m - 2.5 \log{(1 + z)}
\end{equation}
is plotted versus $\log{r}$. The plot of $m^\ast$ versus $\log{r}$ is theoretically linear
with a slope of $5$ since it represents the inverse square law reduction in luminosity in an
Euclidean universe.

The linearity of the plot is the practical reason for plotting $m^\ast$ rather than $m$.
Then, the data can be analyzed by simple linear regression methods. And, since
$m^\ast$ and $m$ only differ by a function of $z$ which is much more accurately
determined than $m$, the error statistics of $m\ast$ and $m$ are very nearly the same.

For supernovae, the Euclidean apparent magnitude $m^\ast = m - 5 \log{(1 + z)}$.
$m\ast$ is increased by $2.5 \log{(1 + z)}$ due to time-dilation (see section~\ref{sec:TD})
of the period of the supernovae light curve. This time-dilation accounts for the observed
anomalous dimming of type Ia supernovae at high $z$.

\subsection{First-Rank Elliptical Galaxies --- Kristian, Sandage \& Westphal}
\label{sec:FRE5}

Two Hubble diagrams for first-rank elliptical galaxies are shown, one from Kristian, Sandage
and Westphal~\cite{ST:KR} and other derived from the surface brightness and effective angular
radii  of first-rank elliptical galaxies (many observers).

The Kristian data will be discussed first. The magnitudes for the first-rank elliptical
galaxies are corrected based on the cluster Abell richness and Bautz-Morgan contrast classes,
the K-correction and galactic absorption.

Figure~\ref{f:fig8}\  shows the a plot of the Euclidean apparent magnitude using aperture corrected data
for the static universe model as described in the next paragraph. Kristian expected a linear
relation between $m$ and $\log(z)$ corresponding to $q = 1$ but actually found that the slope
decreased since the first-rank elliptical galaxies appeared more luminous at higher $z$. The
increased luminosity was attributed to evolution. However, there is another possibility if
the static universe model is correct.

This data was observed before the capability to determine the effective radius was possible.
Consequently, the apparent magnitudes were measured through a fixed angular aperture and then
corrected~\cite{ST:AG}, assuming a $q = 1$ expanding universe model and $H = 50$ km/sec/Mpc,
to the standard physical diameter of a first-rank elliptical galaxy. Since the physical
diameters for the same angular aperture and $z$ are larger in the $q = 1$ expanding universe
model than the static universe model, the apparent magnitudes are less (brighter) than apparent
magnitudes in the static universe model. Therefore, before plotting the data, the
apparent magnitude data was re-corrected to the standard diameter based on the static universe
model. The largest correction was small, only $0.16$ magnitudes less bright.

Then, the following regression relation for the static universe was found
\begin{equation}
    m^\ast = 4.99 \log(r) + 20.62
\end{equation}
with a standard deviation of $0.10$ magnitudes for the slope. Although this data was obtained
in the late 1970's, it is an excellent fit to the static universe model.

\subsection{First-Rank Elliptical Galaxies --- Many Observers}

For the second data set on first-rank elliptical galaxies, the apparent magnitude is
calculated from the surface brightness and the effective angular radius using the equation
\begin{equation}
  m_e = SB - 2.5 \log{(\pi \theta^2)},
\end{equation}
where $\theta$ is the effective angular radius and
\begin{equation}
  m = m_e - 0.75
\end{equation}

For the first-rank elliptical galaxies, the regression equation is
\begin{equation}
   m^\ast = 5.14 \log{(r)} + 21.90.
\end{equation}
The slope is $5.14$ with a standard deviation of $0.16$. This is greater than the
theoretical slope of $5.0$ for a static universe model but is still within one standard
deviation of the theoretical slope.

\begin{figure}
\includegraphics[width=\mypicsize]{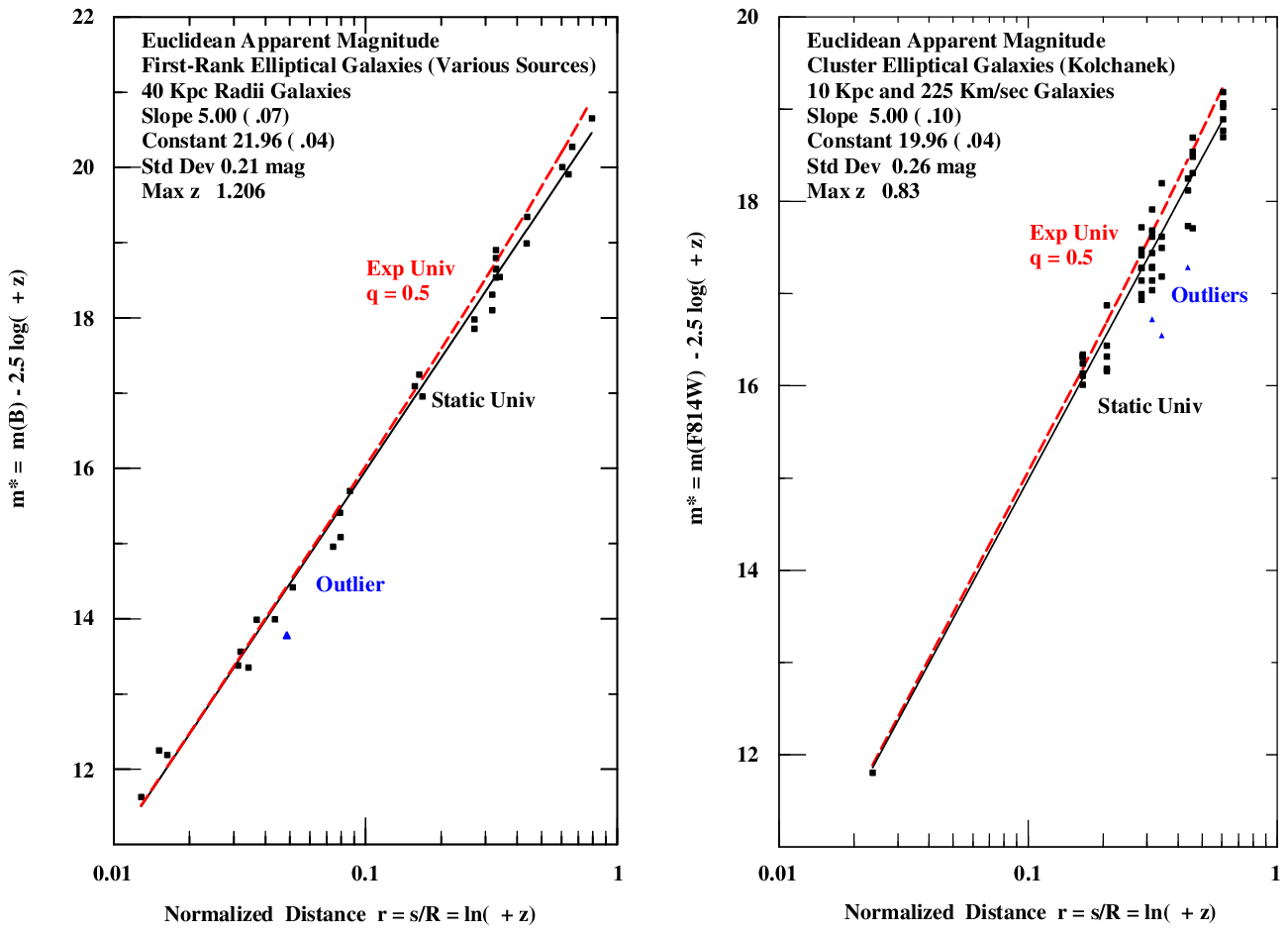}
\caption{Euclidean apparent magnitudes of first-rank and cluster
elliptical galaxies. Both analyzed using the fundamental plane
method. Black lines represent the static universe model observations
with theoretical slope~5. Red dashed lines represent the expanding
universe for $q=0.5$.}
\label{f:fig9}
\end{figure}

However, it is possible to do much better with this data by using the fundamental plane
method. The absolute magnitudes can be referred to a standard first-rank elliptical galaxy
using the fundamental plane relation~\cite{ST:SCH99}
\begin{equation}
   M(B) = -3.33 \log{(R_e)} + \mbox{constant}.
\end{equation}
After carrying through the fundamental plane calculations, the revised regression
equation for the first-rank elliptical galaxies is
\begin{equation}
   m^\ast = 5.00 \log{(r)} + 21.96.
\end{equation}
This regression equation represents a first-rank elliptical galaxy with an effective radius
of $40$ Kpc and is plotted in Figure~\ref{f:fig9}\  (left hand panel). The slope now has a standard
deviation of only $0.07$ magnitudes.

\begin{figure}
\includegraphics[width=\mypicsize]{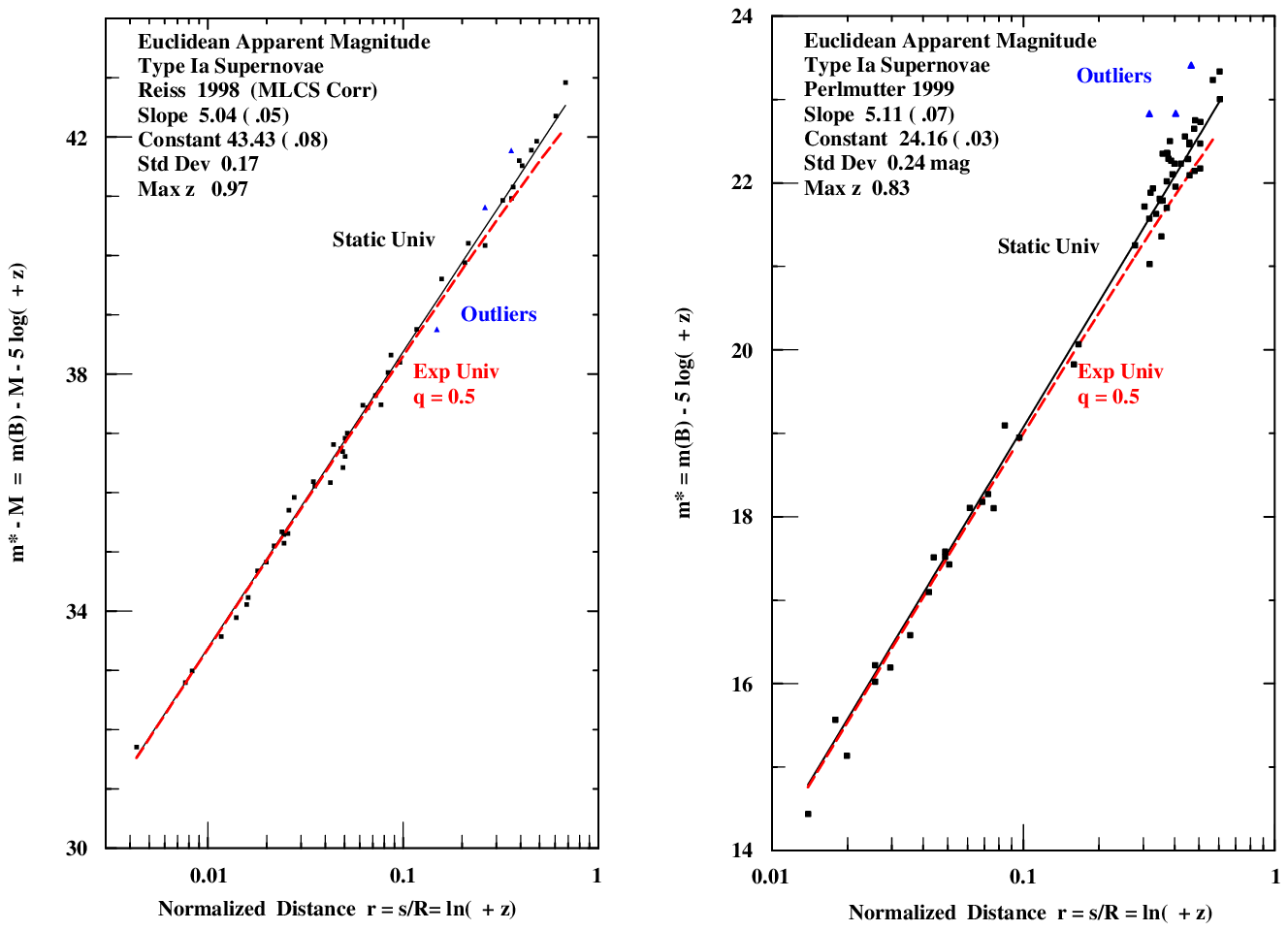}
\caption{Euclidean apparent magnitude for type Ia supernovae. Black
lines with theoretical slope~5.0 represent the static universe model
and red dashed lines represent the theoretical $q=0.5$ expanding
universe model.}
\label{f:fig10}
\end{figure}

\subsection{Cluster Elliptical Galaxies --- Kochanek}
\label{sec:CEG1}

For the cluster elliptical galaxies, the regression equation is
\begin{equation}
   m^\ast = 5.00 \log{(r)} + 19.96.
\end{equation}
This regression equation represents the Euclidean apparent magnitude for a standard
cluster elliptical galaxy with an effective radius of $R_e = 10$ Kpc and a velocity
dispersion $\sigma = 225$ km/sec. Since the standard deviation of the slope is
$0.10$, the static universe model is verified. The regression equation is plotted
in Figure~\ref{f:fig9}\  (right hand panel).

Note: The observationally determined slopes for both the first-rank and cluster elliptical
galaxies are equal to the theoretical slope of $5$ to two decimal points. Although this is a
coincidence (and quite unlikely), it does attest to the precision of the data and the validity
of the static universe hypothesis.

\subsection{Type Ia Supernovae --- Reiss, Perlmutter}
\label{sec:SUP}

Type Ia supernovae observations have been made by two independent supernovae observation
teams. Since for supernovae, $m^\ast = m - 5 \log{(1 + z)}$ is the appropriate quantity to
plot for a linear Hubble relation, the quantity $5 \log{(1 + z)}$ was subtracted from the
maximum apparent magnitudes for the supernovae listed in their papers. The supernovae
observations are plotted in
Figure~\ref{f:fig10}.

The regression equation using the observational data of Reiss~\cite{GR:AR} is
\begin{equation}
   m^\ast - M = m - M - 5 \log{(1 + z)} = 5.04 \log{(r)} + 43.43
\end{equation}
where $M$ is the absolute magnitude. Since the standard deviation of the slope is only $0.05$,
this verifies the static universe model.

The regression equation using the data of Perlmutter~\cite{ST:PERL} is
\begin{equation}
   m^\ast = m - 5 \log{(1 + z)} = 5.11 \log{(r)} + 24.16.
\end{equation}
In this case, the standard deviation is $0.07$, showing that the slope is nearly two standard
deviations from the expected slope of $5$ for the static universe model.

Both of the above regressions include the effect of time-dilation resulting from the variation
of the luminosity of the supernovae over a period of several months or more.

It is important to note that for supernovae, luminosity evolution with $z$ is not expected
because the luminosity only depends on the physics of the explosion. This property of
supernovae is very useful since it can be used to show indirectly that elliptical galaxies also do not evolve with $z$.

To show this, first consider that the good fits of the supernovae observations (Figure~\ref{f:fig10}\ )
confirm the static universe model. Then, it follows logically from the good fits of first-rank
and cluster elliptical galaxies (Figure~\ref{f:fig9}\ ) to the static universe model that first-rank
elliptical galaxies also do not evolve in luminosity with $z$.

\begin{figure}
\includegraphics[width=\mypicsize]{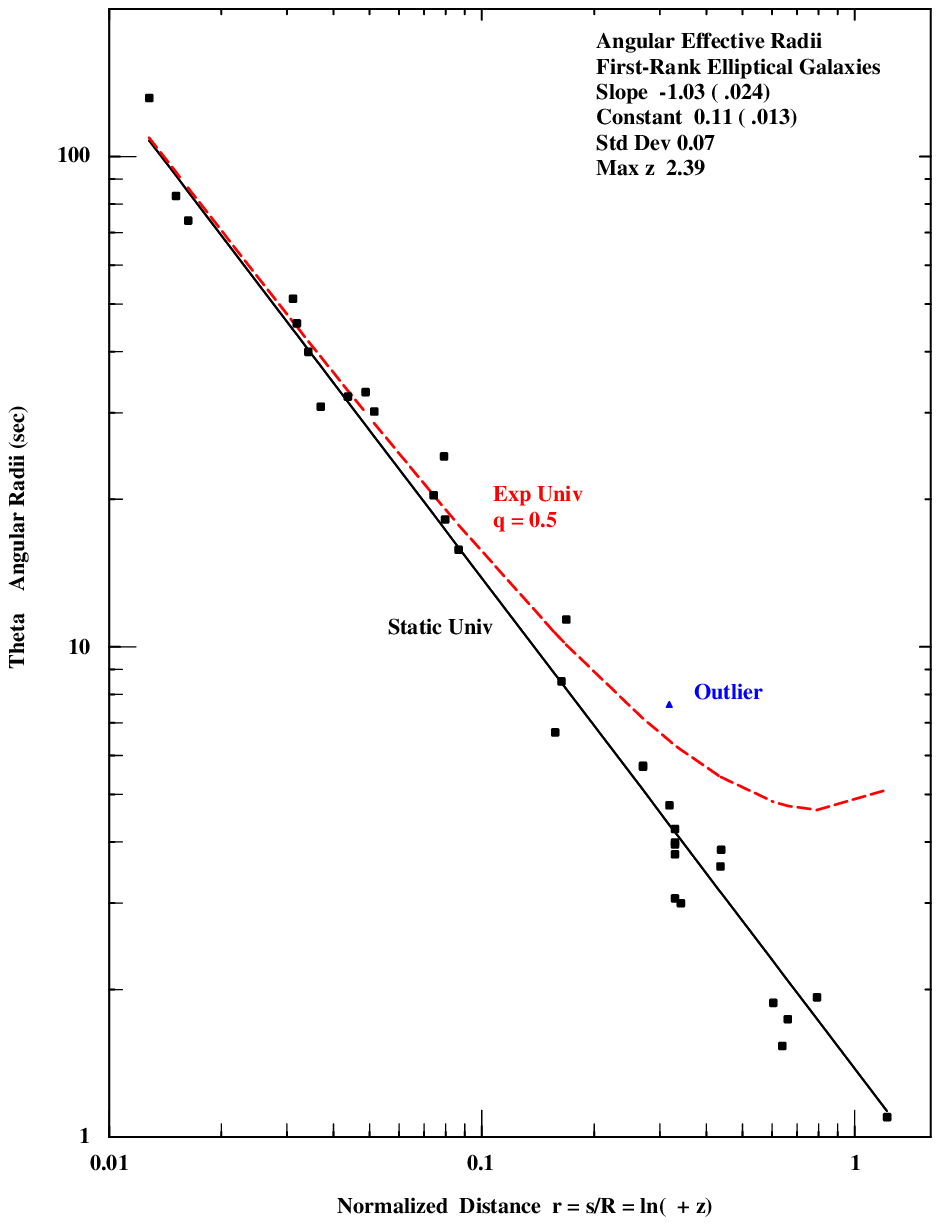}
\caption{Angular radii of first-rank elliptical galaxies. The black
line is the static universe regression line with a slope
of~$-1.03$. The red dashed line represents the $q=0.5$ expanding
universe model with constant radii.}
\label{f:fig11}
\end{figure}

Of course, the non-evolution of elliptical galaxies does appear to contradict current
theoretical studies on the evolution of stars which predict that elliptical galaxies are
brighter at earlier times. Although it is clear that stars evolve, it may be argued,
nevertheless, that other processes exist, for example, the merging of galaxies, that
tend to maintain first-rank elliptical galaxies in an equilibrium state.

Analysis of the supernovae plots shows that supernovae become progressively dimmer than
expected at high z. However, there is no indication of a comparable dimming of first-rank
and cluster elliptical galaxies. (Note that the theoretical $q = 0.5$ expanding universe
curve is on opposite sides of the curves for the supernovae and the first-rank and cluster
elliptical galaxy plots.) Since light propagates identically in the universe for supernovae
and elliptical galaxies, the dimming effect can only be attributed to some feature uniquely
associated with supernovae. This is a strong logical argument for supporting time-dilation
of the supernovae light curve as the correct explanation of the dimming.

\section{Angular Size}
\label{sec:AS}

For the static universe, the theoretical relation between the effective angular
radius, $\theta$, the physical radius, $R_E$ and the normalized distance,
$r = ln(1 + z)$, is given by
\begin{equation}
      \theta = \frac{R_E}{(29.09~r)}.
\end{equation}
If the logarithm's of the variables in the above relation are plotted, the plot is a
straight line with slope $-1$ assuming $R_E$ is constant.
A constant $R_E$ is a critical assumption for first-rank elliptical galaxies but is
confirmed by recent high-z velocity dispersion observations by several observers
of the high z first-rank elliptical galaxies (see Table 3B). Because high-$z$
and low-$z$ first-rank elliptical galaxies have similar velocity dispersions,
they must have similar masses and physical sizes.

\subsection{First-Rank Elliptical Galaxies --- Many Observers}
\label{sec:FRE3}

Angular radii of first-rank elliptical galaxies as shown in Table 3B are plotted in
Figure~\ref{f:fig11}. The regression equation is given by
\begin{equation}
  \log{\theta} = -1.03 \log{(r)} + 0.11.
\end{equation}
Since the standard deviation of the slope is $0.03$, the observations fit the static
universe very well. The theoretical expanding universe model, assuming the effective
physical radii are constant, is also plotted. Because the difference between the models is so
large, the angular size plot strongly supports the static universe model.

This is especially true for the distant radio galaxy observed by Pascarelle~\cite{ST:PASC}
at $z = 2.39$. This galaxy was not used for the surface brightness or apparent magnitude
tests because the surface brightness and the apparent magnitude could not be reliably
determined. However, the angular radius was determined accurately. At the above redshift,
the distance of this galaxy is approximately $24$ billion light years. The age of this
galaxy in the static universe model is much larger than the age of the universe in the
expanding universe models.

\subsection{Double-Lobes of Radio Galaxies --- Nilsson}
\label{sec:DL}

In the 1970's, it was known that plots of the angular diameters versus redshift of the
double lobes of radio galaxies fit the static Euclidean model instead of the expanding
universe model. Nevertheless, a way to save the expanding universe model was found.
Since the power of the double-lobes at high redshift (calculated assuming an expanding
universe) was much greater than at low redshift, it was hypothesized that the physical
diameters of the double-lobes were
inversely proportional to the power. With this hypothesis, the data fit the expanding
universe model. However, this was actually a completely empirical procedure.

In recent years, lower power double-lobes of radio galaxies were
observed at high redshift. Then, it was found that the physical
diameters of the double-lobes of radio galaxies were not inversely
dependent on the power~\cite{GR:BL}. This falsified the hypothesized
inverse power relation.

Here, we specifically check whether the angular diameter data fit the
static universe model. The data for the angular diameters, the redshift
and the power of the double-lobes of radio galaxies are from
Nilsson~\cite{GR:MN}. The large variation in the diameters of the
double-lobes and the generally greater power of the distant radio
galaxies constitute the major problems in the proper analysis of the
data. Most of the variations in the angular diameters at a given
redshift are assumed due to projection effects. A simulation of the
projection effects confirms this assumption.

\begin{figure}
\includegraphics[width=\mypicsize]{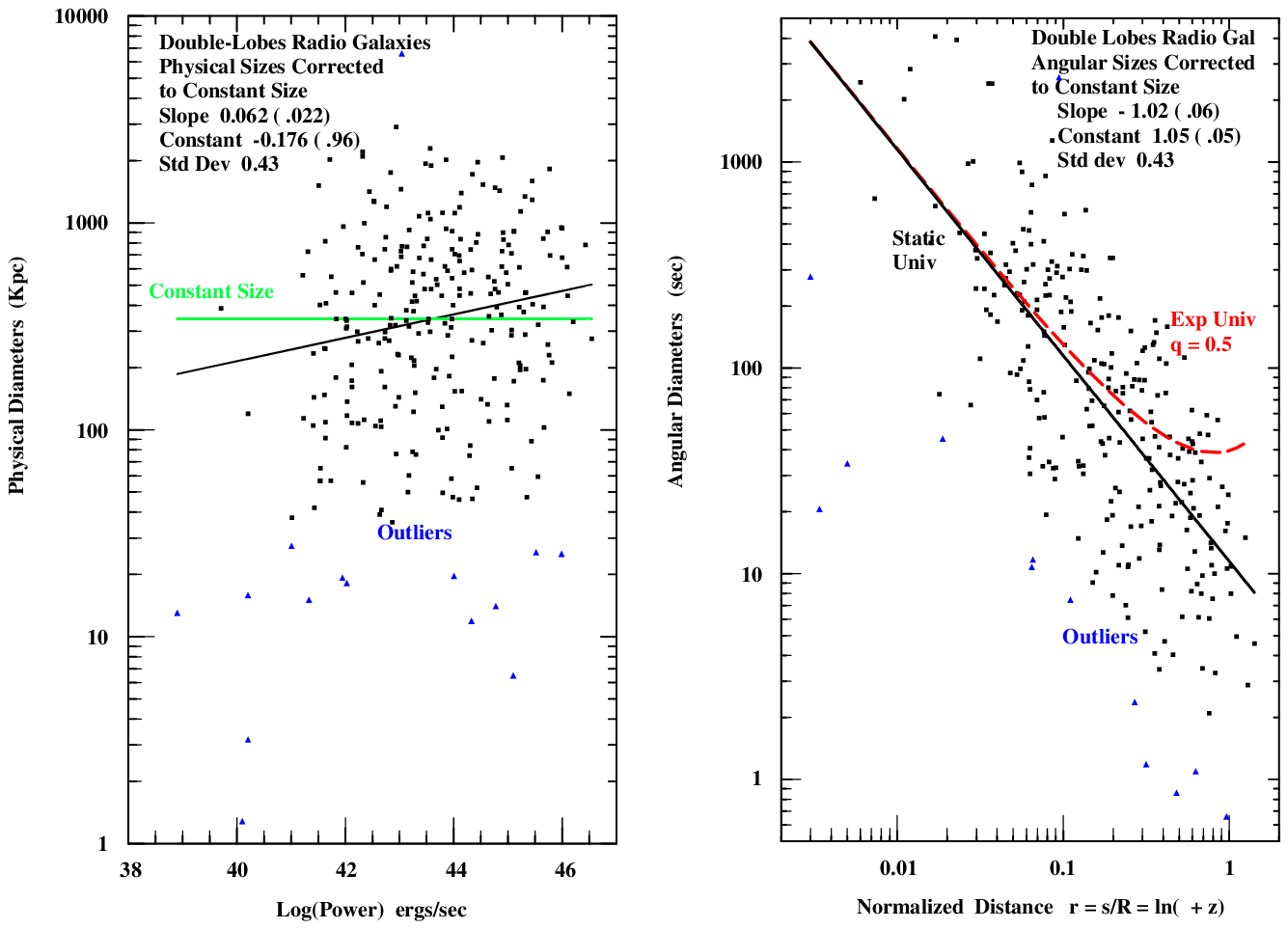}
\caption{Physical and angular radii of double lobes of radio
galaxies. Physical diameters are corrected to a constant size and used
to correct the angular diameters. The green line represents the
corrected constant diameters. The black lines represent the static
universe and the dashed red line the $q=0.5$ expanding universe model.}
\label{f:fig12}
\end{figure}

To check whether the physical diameters may be correlated with the large variation in power,
the physical diameters of the double-lobes are plotted versus the power in Figure~\ref{f:fig12}\ 
(left-hand panel). Both the physical diameters and the power were calculated assuming the
static universe
model. The regression equation between the physical diameters and the power is
\begin{equation}
  \log{(D)} = 0.062 \log{(Power)} - 0.176
\label{ST:DP}
\end{equation}
where $D$ represents the physical diameters of the double-lobes in Kpc. The variation with
power in the physical diameters is, therefore, small compared to the observed range of
physical diameters.

Therefore, before determining the regression between the angular diameters and the distance,
the observed $\log{(D)}$ was corrected as follows: The difference between the average $\log{(D)}$
(corresponding to $343.7$ Kpc) and the $\log{(D)}$ as calculated from the regression,
equation~\ref{ST:DP}, was added to the observed $\log{(D)}$. Then, $\log{(\theta)}$ was
calculated from the corrected $\log{(D)}$. The corrected $\log{(\theta)}$ values were
plotted in Figure~\ref{f:fig12}\  (right-hand panel). The resulting regression is
\begin{equation}
  \log{(\theta)} = -1.02 \log{(r)} + 1.05
\end{equation}
with a standard deviation of 0.06. This result confirms the static universe model.

%
%

\section{Galaxy Counts}
\label{sec:GC}

The counting of galaxies is another way of determining the space-time metric.
Theoretically, the number of galaxies, $N$, brighter than apparent magnitude, $m$,
increase as
\begin{equation}
   \log{(N)} = 0.6~ m + \mbox{constant}
\end{equation}
where $N$ is the cumulative number of galaxies observed to apparent magnitude $m$.
This relation holds to about $m = 17$ and then the slope starts to decrease
because the K-corrections begins to decrease the number of galaxies that are
visible within a given apparent magnitude range. In the last $20$ years, counts
of galaxies have been made to approximately $m = 28$ and compared to calculated counts
assuming various expanding universe models. It has been generally found that the
calculated galaxy counts are considerably less than the observed counts at $m > 20$
unless evolutionary effects are assumed.

More recently, enough redshifts of galaxies have been obtained to make a start on
relating galaxy counts to redshifts.

\subsection{Counts versus Apparent Magnitude --- Tyson}

Differential galaxy counts versus apparent magnitude were calculated for both the static
universe and the $q = 0.5$ expanding universe model. The calculations were based
on the 2dF Survey Schecter luminosity functions~\cite{GR:FL} as shown in Table 5A.
No evolution was assumed in the calculations in either the luminosity or in the
space density of the galaxy types.

The 2dF Survey determined the Schecter luminosity functions for five types of
galaxies --- E/SO, Sab, Sbc, Scd and Sm-Im. For each type of galaxy, K-corrections from
King~\cite{GR:CK} were used to $z = 1.5$. Beyond $z = 1.5$, the K-corrections for
$E/SO$ and $S_{ab}$ galaxies were assumed to increase as $1.25 \log{(1 + z)}$.
For the other type galaxies, the K-corrections were assumed to remain constant
for $z > 1.5$. These K-corrections beyond $z = 1.5$ reflect increases in luminosity
from the ultra-violet moving into the $b_j$ band.

The observations are valid to $z = 4$ where the Lyman break (a
steep reduction in luminosity at 912 angstroms) corresponds to the center
of the observed $b_j$ band at 4500 angstroms. To simulate
the Lyman break, the calculated counts were cutoff at $z = 4$. The calculated
counts were also cutoff at absolute magnitude $-11.0$, a reasonable lower limit for
the luminosity of galaxies. Galactic extinction was set equal to zero.

\begin{figure}
\includegraphics[width=\mypicsize]{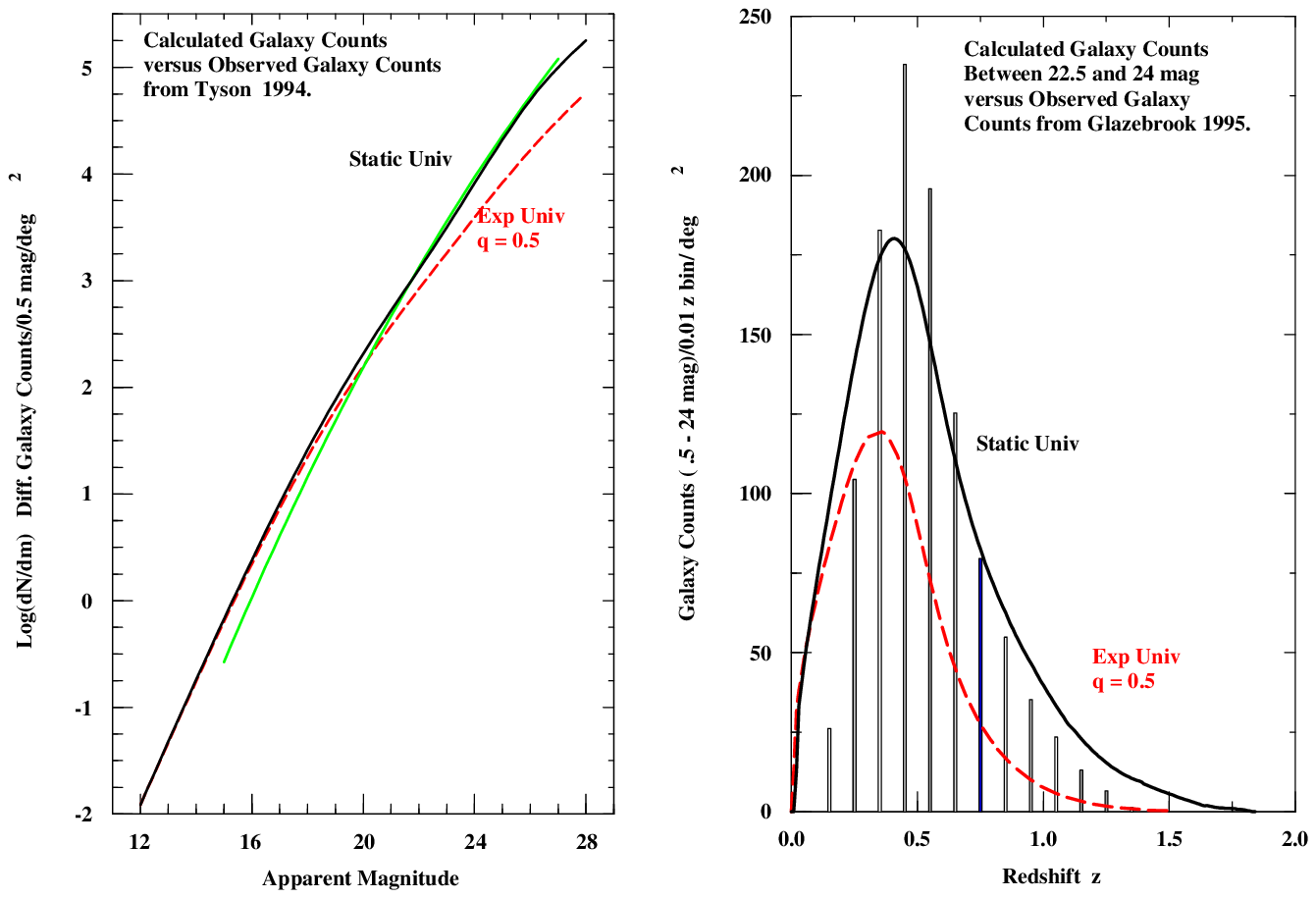}
\caption{The calculated counts for the static universe and the~$q=0.5$
expanding universe are shown by the black line and the dashed red
line, respectively. The green line shows the observed counts by Tyson
between~15 and~27~mag. In the right-hand panel, observed counts versus
redshift are shown as narrow black bars of width~0.01 in~$z$.}
\label{f:fig13}
\end{figure}

The calculations were made using $1/4$ magnitude increments over the range $12$ to $28.5$
magnitudes. It must be said that this required six very large spreadsheets, one for each
type of galaxy. On the other hand, using a spreadsheet makes it very easy to determine
galaxy counts for each type of galaxy and galaxy counts versus redshift.

\begin{table}
\begin{center}
Table 5A \\ \vspace{1mm} 2dF Survey Schecter Luminosity Functions \\
for $H = 100$ km/sec \vspace{1mm}

\begin{tabular}{lccc} \hline \hline

Galaxy Type  & $M(b_j)$* & $\alpha$ & $\phi$* $(10^{-3})$ \\
\hline
E/SO         & -19.61    &  -0.74   &  9.0 \\
Sab          & -19.68    &  -0.86   &  3.9 \\
Sbc          & -19.38    &  -0.99   &  5.3 \\
Scd          & -19.00    &  -1.21   &  6.5 \\
Sm/Im        & -19.02    &  -1.73   &  2.1 \\
\hline \hline

\end{tabular}

\end{center}
\end{table}

The calculated and observed differential galaxy counts are plotted in
Figure~\ref{f:fig13}\  (left hand panel). The observed differential galaxy counts
from Tyson~\cite{GR:JT} are represented by the green line, valid between
$m = 15$ and $27$ in the $b_j$ band. However, note that to $m = 20$, the observed
counts are less than the calculated counts since Tyson initially picked
out areas of sky apparently devoid of galaxies to roughly $m = 19$.
But, from $m = 20$ to $27$ magnitudes,
the observed and calculated counts for the static universe are equal.

On the other hand, the calculated differential galaxy counts based on the
expanding universe model with $q = 0.5$ with no evolution are significantly
less than the observed counts.

Since the differential galaxy counts in the static universe model are very
closely the same as the observed galaxy counts, this is good evidence that medium and
high redshift galaxies have the same luminosity distribution and space density
as local galaxies. Consequently, the observations both confirm the static universe
model and the PCP hypothesis of an equilibrium (no evolution) universe.

\subsection{Counts versus Redshift --- Glazebrook}

Figure~\ref{f:fig13}\  (right hand panel) shows the calculated number of galaxies within the apparent
magnitude range $22.5 < m < 24.0$ for the static and expanding universe models versus the
observed number of galaxies (scaled to a square degree) in redshift bins $0.01$ z wide. The
observed counts are from Glazebrook~\cite{ST:GL}. Glazebrook observed 73 galaxies in a $73\%$
complete sample within $7$ separate regions with a total angular area of $38.01$ sq. arcmin.
To compare the observed counts with the calculated counts, the number of counts in
Glazebrook's paper were multiplied by a scale factor of $13.05$ to conform to the same scale
as the calculated counts. The scale factor also accounts for the sample's $73\%$ completeness
assuming that the additional galaxies with no measured redshifts have the same redshift
distribution. A reasonable agreement with the static universe calculations is observed,
again confirming the static universe model.

Since both sets of observations agree with the calculated counts, it was thought useful to
also show calculated cumulative galaxy counts by galaxy type to apparent magnitudes $24$
and $28$ for the range of redshifts from $z = 0.25$ to $z = 4$. These cumulative counts
are shown in Tables 5B and 5C. The analysis of the data in the tables shows several
important features: First, few if any E/SO galaxies can be observed at $m < 24$ beyond $z = 0.75$. Other type galaxies can be observed to, at most, $z = 1.5$. For $m < 28$, most E/SO galaxies can be observed only to approximately $z = 2$. Very few can be observed to $z = 3$ or $z = 4$. Second, the large numbers of ``blue'' galaxies at high redshift appear to be Sm-Im. These
exist in approximately the number observed by Tyson at $m> 27$.

\begin{table}
\begin{center}
Table 5B \\
\vspace{1mm}
Cumulative Galaxy Counts to Apparent Magnitude $24$\\

(Logarithm of Galaxy Counts Shown Below)\\

\vspace{1mm}

\begin{tabular}{lccccccc}
\hline \hline
z        & 0.25 & 0.50 & 0.75 & 1.0  & 1.5  & 2.0  & 3.0  \\
r        & 0.22 & 0.41 & 0.56 & 0.69 & 0.92 & 0.92 & 1.10 \\
\hline
E/SO    & 2.76 & 3.29 & 3.38 & 3.38 & 3.38 & 3.38 & 3.38 \\
Sab     & 2.52 & 3.08 & 3.28 & 3.35 & 3.38 & 3.38 & 3.38 \\
Sbc     & 2.52 & 3.11 & 3.33 & 3.44 & 3.52 & 3.53 & 3.53 \\
Scd     & 3.00 & 3.43 & 3.56 & 3.61 & 3.64 & 3.65 & 3.65 \\
Sm-Im   & 3.23 & 3.49 & 3.59 & 3.63 & 3.67 & 3.68 & 3.68 \\
\hline
Totals  & 3.60 & 4.01 & 4.14 & 4.20 & 4.23 & 4.24 & 4.24 \\
\hline   \hline

\end{tabular}

\end{center}
\end{table}

\begin{table}
\begin{center}
Table 5C \\
\vspace{1 mm}
Cumulative Galaxy Counts to Apparent Magnitude $28$\\

(Logarithm of Galaxy Counts Shown Below)\\

\vspace{1 mm}

\begin{tabular}{lcccccccc}
\hline \hline
z       & 0.25 & 0.50 & 0.75 & 1.0  & 1.5  & 2.0  & 3.0  & 4.0 \\
r       & 0.22 & 0.41 & 0.56 & 0.69 & 0.92 & 1.10 & 1.39 & 1.61 \\
\hline
E/SO    & 2.91 & 3.62 & 3.96 & 4.13 & 4.28 & 4.33 & 4.36 & 4.37 \\
Sab     & 2.71 & 3.42 & 3.77 & 4.00 & 4.27 & 4.42 & 4.59 & 4.68\\
Sbc     & 2.71 & 3.43 & 3.79 & 4.03 & 4.32 & 4.49 & 4.70 & 4.81\\
Scd     & 3.53 & 4.11 & 4.39 & 4.57 & 4.73 & 4.93 & 5.07 & 5.14\\
Sm-Im   & 4.25 & 4.60 & 4.91 & 4.87 & 5.01 & 5.09 & 5.17 & 5.20\\
\hline
Totals  & 4.36 & 4.79 & 5.10 & 5.16 & 5.33 & 5.46 & 5.58 & 5.64\\
\hline   \hline

\end{tabular}

\end{center}
\end{table}

\section{Summary}
\label{sec:SUM}

Because the static universe hypothesis is a simple and logical
deduction from the PCP and the observational data amply confirms the
deductions of the static universe hypothesis, I conclude that the
universe is static and in an equilibrium state. I also find that the
new gravitational theory is confirmed by the cosmological observations.

\section{Acknowledgments}

 I acknowledge the excellent work of the observers referenced in this paper. Without their precise observations, I could not have completed this work. As is well known, the scientific method requires that new theories must be verified by observations.


\pagebreak

\section{List of Figures}

\begin{description}
\item[New Physics]
  \item[  Figure~\ref{f:fig1}\ ] Measured Vertical Gravity Variations During Solar Eclipse
  \item[  Figure~\ref{f:fig2}\ ] Average Density of Particles in the Universe
\item[Redshift Process]
  \item[  Figure~\ref{f:fig3}\ ] Simplified Model of Interacting Particles in the Universe
\item[Surface Brightness]
  \item[  Figure~\ref{f:fig4}\ ] First-Rank Elliptical Galaxies (Many Observers)
  \item[  Figure~\ref{f:fig5}\ ] Distribution of Parameters
  \item[  Figure~\ref{f:fig6}\ ] Cluster Elliptical Galaxies
  \item[  Figure~\ref{f:fig7}\ ] Distribution of Parameters
  \item[Euclidean Apparent Magnitude]
  \item[  Figure~\ref{f:fig8}\ ] First-Rank Elliptical Galaxies (Kristian, Sandage \& Westphal)
  \item[  Figure~\ref{f:fig9}\ ] First-Rank (Many Observers) and Cluster Elliptical Galaxies
  \item[  Figure~\ref{f:fig10}\ ] Type Ia Supernovae
\item[Angular Size vs Distance]
  \item[  Figure~\ref{f:fig11}\ ] Angular Radii of First-Rank Elliptical Galaxies
  \item[  Figure~\ref{f:fig12}\ ] Angular Diameters Double Lobes of Radio Galaxies
\item[Galaxy Counts]
  \item[  Figure~\ref{f:fig13}\ ] Galaxy Counts vs Apparent Magnitude and Redshift
\end{description}

\pagebreak


\section{Appendix --- Simulation of Hubble Redshift}

This is the basic program I used to calculate the change in energy due to moving a mass particle or a photon a short distance, $s$, in the universe.

\begin{verbatim}

'File Redshift.BAS (Written in MS-DOS QBasic)
'Hubble Redshift Simulation for a Photon or Mass Particle).

'Problem Description:
'To simulate the interaction energy of a particle with all the
'particles in the universe before and after moving the particle
'a small distance, s, to the right along the $x$-axis.

'The Particle's initial position is denoted by Pos1. The final
'position, is denoted by Pos2. Energy inputs from particles
'originate in regions RA, RB and RC. The interaction is of the
'form exp(-r1)/r1 for the particle in position 1 and
'exp(-r2)/r2 for region A, position 2 and exp(-r3)/r3 for
'region C, position 2.

'In RA1, the particle distance is r1 = SQR(x^2 + y^2 + z^2) away 
'from particles in the left hand quadrant. Normalized, the effect
'is one-half of the interaction with all the particles in the
'universe.

'In RA2, the moved particle is further away from the particles on
'the left in the universe. Then, r2 = SQR((x + s)^2 + y^2 + z^2).

'In RC2, the moved particle is closer to the particles on the
'right in the universe. Then, r3 = SQR((x - s)^2 + y^2 = z^2).

'The integrations are limited to equal or less than maxs!.

'Results are written to the file "Results.BAS".


'Begin Program

OPEN "RESULTS.BAS" for OUTPUT as #1

'Set Parameters
progi% = 352   'Program Iterations
maxs! = 7.0    'Maximum Radius
ints! = 0.02   'Integration Step
intn% = 3      'Number of Integr Steps Particle Moved to Right
s! = intn%*ints!           'Distance Particle Moved
pi! = 3.1415926
c! = 4/(4*pi!)*ints!^3     'For Half-Sphere Normalization,
'results correspond to interactions with particles in half
'the universe.

'Initializations
ERA1# = 0
ERA2# = 0
ERB1# = 0
ERC1# = 0

'Main Program

PRINT "REDSHIFT.BAS:  "; "Energy Absorption of a Particle"
PRINT "               "; "at Two Positions, P1 and P2."
PRINT " "
FOR i% = 0 TO progi%
  x! = i% * ints!

  FOR j% = 0 TO progi%
    y! = j% * ints!

    FOR k% = 0 TO progi%
      z! = k% * ints!

      Temp! = y!^2 + z!^2
      r1# = SQR(x!^2 + Temp!)        'Dist for Particle 1A
      r2# = SQR((x! + s!)^2 + Temp!) 'Dist for Particle 2A
      r3# = SQR((x! - s!)^2 + Temp!) 'Dist for particle 2C
      IF r1# = 0 THEN r1# = 20
      IF r3# = 0 THEN r3# = 20

'Region A for Particle 1
      IF r1# <= maxs! THEN                 'Sum to Max Radius
      ERA1# = ERA1# + EXP(-r1#)/r1#
      IF i% = 0 THEN ERA1# = ERA1# - EXP(-r1#)/(2*r1#)
      IF j% = 0 THEN ERA1# = ERA1# - EXP(-r1#)/(2*r1#)
      IF k% = 0 THEN ERA1# = ERA1# - EXP(-r1#)/(2*r1#)
      END IF

'Region A for Particle 2
      IF r2# <= maxs! THEN                 'Sum to Max Radius
      IF i% >= intn% THEN
         ERA2# = ERA2# + EXP(-r2#)/r2#
         IF i% = intn% THEN ERA2# = ERA2# - EXP(-r2#)/(2*r2#)
         IF j% = 0 THEN ERA2# = ERA2# - EXP(-r2#)/(2*r2#)
         IF k% = 0 THEN ERA2# = ERA2# - EXP(-r2#)/(2*r2#)
      END IF
      END IF

'Region B for Particle 1 (Same Result for Particle 2)
      IF r1# <= maxs! THEN                'Sum to max radius
      IF i% <= intn% THEN
        ERB1# = ERB1# + EXP(-r1#)/r1#
        IF i% = 0 THEN ERB1# = ERB1# - EXP(-r1#)/(2*r1#)
        IF i% = intn% THEN ERB1# = ERB1# - EXP(-r1#)/(2*r1#)
        IF j% = 0 THEN ERB1# = ERB1# - EXP(-r1#)/(2*r1#)
        IF k% = 0 THEN ERB1# = ERB1# - EXP(-r1#)/(2*r1#)
      END IF
      END IF

'Region C for Particle 2
      IF r3# <= maxs! THEN                'Sum to Max Radius
      IF i% >= intn% THEN
         ERC1# = ERC1# + EXP(-r3#)/r3#
         IF i% = intn% THEN ERC1# = ERC1# - EXP(-r3#)/(2*r3#)
         IF j% = 0 THEN ERC1# = ERC1# - EXP(-r3#)/(2*r3#)
         IF k% = 0 THEN ERC1# = ERC1# - EXP(-r3#)/(2*r3#)
      END IF
      END IF

'Output to Screen
      prcheck% = INT(SQR((maxr#^2/2)))
      IF J% = prcheck%  and k% = prcheck% THEN
      ERA1! = ERA1#*c!
      ERA2! = ERA2#*c!
      ERB1! = ERB1#*c!
      ERC1! = ERC1#*c!
      DiffA1A2! = ERA1! - ERA2!
      DiffA1C1! = ERA1! - ERC1!
      PRINT x!; ERA1!; ERA2!; ERB1!; ERC1!; DiffA1A2!; DiffA1C1!
      END IF

      NEXT k%
   NEXT j%
 NEXT i%

'Output to File
ERA1! = ERA1#*c!
ERA2! = ERA2#*c!
ERB1! = ERB1#*c!
ERC1! = ERC1#*c!
DiffA1A2! = ERA1! - ERA2!
DiffA1C1! = ERA1! - ERC1!
Print #1, " "
Print #1, "Program REDSHIFT.BAS:    "; "Energy Absorption of Particle"
Print #1, " "
Print #1, "Parameters: ";
Print #1, " "
Print #1, "Program Iterations (progi%) = "; progi%
Print #1, "Maximum Radius (maxr#)      = "; maxs!
Print #1, "Integration Step (ints!)    = "; ints!
Print #1, "Particle Steps (intn%)      = "; intn%
Print #1, " "
Print #1, "Particle Position (s!)      = "; s!
Print #1, " "
Print #1, "Results"
Print #1, " "
Print #1, "EnergyRA1 (ERA1#) =: "; ERA1!
Print #1, "EnergyRA2 (ERA2#) =: "; ERA2!
Print #1, "EnergyRB1 (ERB1#) =: "; ERB1!
Print #1, "EnergyRC1 (ERC1#) =: "; ERC1!
Print #1, " "
Print #1, "Difference A1A2 (DiffA1A2#) = "; DiffA1A2!
Print #1, "Difference A1C1 (DiffA1C1#) = "; DiffA1C1!
Print #1, " "
Print #1, "End Results"

CLOSE #1
END

The output of the program for one set of parameters is
shown below: The main result is Diff A1A2 = 0.00987
which should equal (s/R)E = 0.01E.

Program REDSHIFT.BAS:  Energy Absorption of Particle
 
Parameters:  
Program Iterations (progi%) =  1002 
Maximum Radius (maxr#)      =  10 
Integration Step (ints!)    =  0.01 
Particle Steps (intn%)      =  1 
 
Particle Position (s!)      =  0.01 
 
Results
 
EnergyRA1 (ERA1#) =:  .499629 
EnergyRA2 (ERA2#) =:  .4897593 
EnergyRB1 (ERB1#) =:  4.799089E-3 
EnergyRC1 (ERC1#) =:  .499629 
 
Difference A1A2 (DiffA1A2#) =  9.869665E-3 
Difference A1C1 (DiffA1C1#) =  0 
 
End Results

\end{verbatim}

\end{document}